\documentclass[12pt]{article}
\usepackage[dvips]{graphicx}

\newcommand{\ba}{\begin{eqnarray}}
\newcommand{\ea}{\end{eqnarray}}
\newcommand{\baa}{\begin{array}}
\newcommand{\eaa}{\end{array}}

\renewcommand{\thesection}{\arabic{section}}

\newcommand{\beq}{\begin{equation}}
\newcommand{\eeq}{\end{equation}}

\newcommand{\Cg}{{\mathcal C}_{\rm g}}
\newcommand{\Tr}{{\mathrm{Tr}}}

\newcommand{\hps}{h_{\phi}}
\newcommand{\hsc}{h_{\sigma}}
\newcommand{\minimumpoint}{
\begin{array}{l}
\scriptstyle\phi=0 \\[-2mm]\scriptstyle \sigma=0\\[-2mm]\scriptstyle \chi=\chi_c
\end{array}
}
\newcommand{\sigmaI}{\sigma_{\rm I}}
\newcommand{\sigmaII}{\sigma_{\rm II}}
\newcommand{\sigmaIII}{\sigma_{\rm III}}
\newcommand{\sigmaIV}{\sigma_{\rm IV}}
\newcommand{\sigmaV}{\sigma_{\rm V}}
\newcommand{\comment}[1]{}

\newcommand{\jj}[3]{ {j}_{{\rm #1}, #2}^{#3} }
\newcommand{\FF}[3]{ {F}_{{\rm #1}, #2}^{#3} }
\newcommand{\JJ}[3]{{\cal J}^{\Lambda}_{#1,#2}[#3]}
\newcommand{\Decay}[2]{\Gamma_{#1}^{{\rm #2}}}
\newcommand{\mev}{\;{\rm MeV}}
\newcommand{\mbar}{{\overline{m}}}
\newcommand{\vev}[1]{\left\langle#1\right\rangle_0}

\begin{document}

\title{ Ground and excited scalar isoscalar meson
states in a $U(3)\times U(3)$
quark model with a glueball.}
\author{M. K. Volkov, V. L. Yudichev\\
\itshape Bogoliubov Laboratory of Theoretical Physics,\\
\itshape Joint Institute for Nuclear Research,\\
\itshape 141980 Dubna, Russia}
\date{}
\maketitle

\begin{abstract}
Ground and first radially excited scalar isoscalar
meson states and a scalar glueball are described in a nonlocal
$U(3)\times U(3)$ quark model.
The glueball is introduced
into the effective meson Lagrangian by means of the dilaton
model on the base of the scale invariance of the meson
Lagrangian.
The scale invariance breaking by current quark masses and gluon
anomalies is taken into account.
The glueball anomalies turn out to be important for
the description of the glueball-quarkonia mixing.
The masses of five scalar isoscalar meson states and their
strong decay widths are calculated. The state $f_0(1500)$
is shown to be composed mostly of the scalar glueball.
\end{abstract}


\newpage
\setcounter{page}{1}

\section{Introduction}
In our recent papers \cite{IJMPA1999,YF,ECHAJA2000},
it was shown that the experimentally observed scalar
meson states lying in the mass interval from 0.4 to
1.7 GeV \cite{PDG} can be interpreted
as two nonets of scalar quarkonia:
the ground state nonet (with masses below 1 GeV) and
the nonet of their first radial excitations  (heavier than 1 GeV).
Meanwhile, it is established from experiment
that another scalar isoscalar meson state exists
in this mass interval \cite{PDG}.
It is used to be associated with the scalar glueball.
The most probable candidates for the glueball
are the states $f_0(1500)$ and $f_0(1710)$
\cite{Anisovich,Jaminon,LeeWeingarten}.
In \cite{IJMPA1999,YF,ECHAJA2000}, we came to the conclusion
that $f_0(1710)$ is rather a quarkonium, while
$f_0(1500)$ is a glueball.
This is in agreement
with the results given in \cite{Anisovich}.
However, to make the final decision, one should
introduce the glueball into the effective meson Lagrangian.
Our present paper is devoted to solution of this problem.

A nonlocal version of the $U(3)\times U(3)$ chiral quark model
with the local 't Hooft interaction
\cite{IJMPA1999,YF,ECHAJA2000} was used to
describe the meson nonets mentioned above.
The nonlocality was introduced there by means of form factors
in quark currents. This allowed us to describe
the nonet of first radial excitations
\cite{ECHAJA2000,VolkovWeiss,Volk1997}.
The form factors were chosen so that
they allow to satisfy the low-energy
theorems in the chiral limit
and keep gap-equations in a form
derived from the standard Nambu--Jona-Lasinio (NJL) model.
In the momentum space, these form factors
are expressed through first degree polynomials
depending on the momentum squared and have
a Lorentz-covariant form.
The masses and decays of the ground and radially excited
nonets of scalar, pseudoscalar, and vector mesons
were described in the framework
of this model \cite{IJMPA1999,YF,ECHAJA2000}.
However, we did not consider the glueball.
Here we suggest an extended version of the non-local
$U(3)\times U(3)$ quark model that gives
a description of the scalar glueball as well as
the ground and radially excited scalar quarkonia nonets.

A common method of introducing the glueball into
the effective meson Lagrangian is to take advantage of
the dilaton model.
The dilaton model was used by many authors
\cite{Jaminon,Kusaka,Andrianov} for this purpose.
These models are
based on the approximate scale invariance of the
effective meson Lagrangian, which is in accordance with
the QCD Lagrangian scale invariance if the current quark masses
are equal to zero.
As in QCD, in the effective meson Lagrangian,
the terms with current quark masses also break scale invariance.
Moreover, the scale invariance is  broken by terms induced
by gluon anomalies, which is also in accordance with QCD.
All the terms that break scale invariance
turn out to be important for the description of
the quarkonia-glueball mixing and, as a consequence, have
a noticeable effect on the strong decay modes of scalar mesons.

In papers \cite{Acta,EPJA2000,YF2000}, we constructed a model
describing only the ground scalar isoscalar meson states
and a glueball. It was shown that the state $f_0(1500)$ is
rather the scalar glueball than $f_0(1710)$.
We described its decays
in satisfactory agreement with available experimental data.
We also found that the terms connected with gluon anomalies
determine the most of quarkonia-glueball mixing.

Here, we extend our model to describe both the ground and
radially excited scalar isoscalar quarkonia as well as
the scalar glueball state. Thereby, we obtain the complete
description of 19 scalar meson states within the mass interval
from 0.4 to 1.7 GeV. Our approach and results noticeably
differ from those given in  \cite{Jaminon,other}. \tolerance=7000
Moreover, for the first time, we succeeded to describe the
nature of  all 19 scalar  meson states.

Insofar as we cannot expect that the chiral symmetry
can determine the properties of so heavy particles well enough,
we claim here only qualitative
agreement of our results with experiment.
Only isoscalar states are considered.
Concerning the isovector and strange mesons,
the introduction of the scalar glueball
changes little the results obtained for them
in \cite{IJMPA1999,YF,ECHAJA2000}.

The structure of our paper is following. In section 2,
a nonlocal chiral quark model of the NJL type with the local six-quark
't Hooft interaction is bosonized to construct an
effective meson Lagrangian.
In section~3, the meson Lagrangian is
extended by introducing a scalar glueball as a dilaton
on the base of scale invariance. The gap equations,
the divergence of the dilatation current and
quadratic terms of the effective meson Lagrangian are derived
in sect.~4. There, we also diagonalize  quadratic
terms.  Numerical
estimates of the model parameters are given in sect.~5.
In section 6, the widths for the main modes of strong decays of
scalar isoscalar mesons are calculated. The discussion over
the obtained results is given in sect.~7.
A detailed description of how to calculate the
quark loop contribution to the width of strong
decays of scalar mesons is given Appendix A.

\section{$U(3)\times U(3)$ Lagrangian for quarkonia}

We start from an
effective $U(3)\times U(3)$ quark Lagrangian
of the following form (see \cite{IJMPA1999,YF,ECHAJA2000}):
\ba
L& =& L_{\rm free}+L_{\rm NJL}+L_{\rm tH},\label{Ldet}\\
L_{\rm free}&=&{\bar q}(i{\hat \partial} - m^0)q\\
L_{\rm NJL}&=&
 {G\over 2}\sum_{i=1}^{N}\sum_{a=1}^9
[(\jj{S}{i}{a})^2 +
(\jj{P}{i}{a})^2]\label{njl}\\
L_{\rm tH}&=&- K \left\{ {\det}[{\bar q}(1+\gamma_5)q]+{\det}[{\bar
q}(1-\gamma_5)q]
\right\},
\ea
where $L_{\rm free}$ is the free quark Lagrangian with
$q$ and $\bar q$  being $u$, $d$, or $s$ quark fields;
$m^0$ is a current quark mass matrix with
diagonal elements: $m^0_{\rm u}$, $m^0_d$, $m^0_{\rm s}$ $(m^0_{\rm u} \approx m^0_d)$.
The term $L_{\rm NJL}$ contains nonlocal four-quark vertices
of the Nambu--Jona-Lasinio (NJL) type which have the current-to-current
form. The quark currents are defined in accordance with
\cite{IJMPA1999,YF,ECHAJA2000,VolkovWeiss,Volk1997}:
\beq
\jj{S(P)}{i}{a}(x)=
\int\!d^4x_1d^4x_2 \bar q(x_1)\FF{S(P)}{i}{a}(x;x_1,x_2)q(x_2),
\label{j}
\eeq
where the subscript S is for scalar and P for the pseudoscalar currents.
The term $L_{\rm tH}$ is the six-quark 't~Hooft interaction
which is supposed to be local, so no form factor is introduced
in $L_{\rm tH}$.

Currents (\ref{j}) are nonlocal due to the nonlocal quark vertex functions
$\FF{S(P)}{i}{a}$. This way of introducing nonlocality allows
to consider radially excited meson states, which is impossible
in the standard NJL model. In general, the number of radial
excitations $N$ is infinite, but we restrict our-selves with
$N=2$, leaving only the ground and first radially excited states,
because extending this model by involving
more heavier particles
is not valid for this class of models.

Let us define the quark vertex functions in the momentum space.
\ba
 &&\FF{S(P)}{i}{a}(x;x_1,x_2)=\nonumber\\
&&\quad= \int\!\frac{d^4P}{(2\pi)^4}
    \frac{d^4 k}{(2\pi)^4}
    \exp \frac{i}{2}
    \biggl(
    (P+k)(x-x_1)\nonumber\\
&&\quad+(P-k)(x-x_2)\biggr)\FF{S(P)}{i}{a}(k|P),
\ea
where $P$ is the total momentum of a meson and
$k$ is the relative momentum of quarks inside
the meson.
As it was mentioned in the Introduction,
here we follow papers
\cite{IJMPA1999,YF,ECHAJA2000,VolkovWeiss,Volk1997},
where the functions $\FF{S(P)}{i}{a}(k|P)$ are chosen
in the momentum space as follows:
\beq
\FF{S}{i}{a}(k|P)=\tau_a f_i^a(k_\perp),\quad
\FF{P}{i}{a}(k|P)=i\gamma_5\tau_a f_i^a(k_\perp),
\eeq
and the form factors $f_i^a$, $(i=1,2)$ are
\beq
f_1^a(k_\perp)=1,\quad f_2^a(k_\perp)=c_a(1+d_a |k_\perp|^2).
\label{formfactors}
\eeq
The form factors depend on the transverse relative momentum
of the quarks:
\beq
k_\perp=k-\frac{P\cdot k}{P^2}P.
\eeq
In the rest frame of a meson, the vector $k_\perp$ equals
$(0,\vec{k})$, thereby the form factors can be considered
as functions of 3-dimensional momentum. Further calculations
will be carried out in this particular frame.
The matrices $\tau_a$ are related to the Gell-Mann
$\lambda_a$ matrices as follows:
\ba
&&{\tau}_a={\lambda}_a
~~~ (a=1,...,7),~~~\tau_8 = ({\sqrt 2} \lambda_0 + \lambda_8)/{\sqrt
3},\nonumber\\ &&\tau_9 = (-\lambda_0 + {\sqrt 2}\lambda_8)/{\sqrt 3}.
\ea
Here   $\lambda_0 = {\sqrt{2/ 3}}$~{\bf 1}, with {\bf
1} being the unit matrix.

The first form factor is equal to unit. This corresponds to
the standard NJL model which we obtain in the case $N=1$.
Let us note that, with the introduction of form factors
for  radially excited states,
new parameters $c_a$ and $d_a$ appear in the model.
This requires additional data to fix them.
The internal (slope) parameter is fixed theoretically
(see eq. (\ref{zerotadpole}) in sect.~4), while
the external parameter $c_a$ is determined from
the mass spectrum  of pseudoscalar mesons.

It is convenient to use an equivalent form of
Lagrangian (\ref{Ldet}) containing only four-quark vertices
whose interaction constants take account of
the 't Hooft interaction. Using the method described in
\cite{EPJA2000} and
\cite{Cimen_99,Vogl_91,Kleva_92}, we obtain
\ba
&&L = {\bar q}(i{\hat \partial} -
\mbar^0)q \nonumber\\
&&\qquad+{1\over 2}\sum_{a, b=1}^9\left[G_{ab}^{(-)}
\jj{S}{1}{a}\jj{S}{1}{b} +
 G_{ab}^{(+)}\jj{P}{1}{a}\jj{P}{1}{b}
\right]\nonumber\\
&&\qquad+\frac{G}{2}\sum_{a=1}^{9}
\left[\jj{S}{2}{a}\jj{S}{2}{a}+\jj{P}{2}{a}\jj{P}{2}{a}\right]
, \label{LGus}
\ea
where
\ba
&&G_{11}^{(\pm)}=G_{22}^{(\pm)}=G_{33}^{(\pm)}= G \pm
4Km_{\rm s}I^\Lambda_1(m_{\rm s}), \nonumber \\
&&G_{44}^{(\pm)}=G_{55}^{(\pm)}=G_{66}^{(\pm)}=G_{77}^{(\pm)}= G \pm
4Km_{\rm u}I^\Lambda_1(m_{\rm u}), \nonumber \\
&&G_{88}^{(\pm)}= G \mp
4Km_{\rm s}I^\Lambda_1(m_{\rm s}),\quad G_{99}^{(\pm)}= G,\nonumber\\
&&G_{89}^{(\pm)}=G_{98}^{(\pm)}= \pm 4{\sqrt
2}Km_{\rm u}I^\Lambda_1(m_{\rm u}),\nonumber\\&&G_{ab}^{(\pm)}=0\quad (a\not=b; \quad
a,b=1,\dots,7),\nonumber\\
&&G_{a8}^{(\pm)}=G_{a9}^{(\pm)}=G_{8a}^{(\pm)}=G_{9a}^{(\pm)}=0\quad (a=1,\dots, 7),
\label{DefG}
\ea
and $\mbar^0$ is a diagonal matrix composed of modified current quark masses:
\ba
    \overline{m}^0_{\rm u}&=&m^0_{\rm u}- 32 K m_{\rm u} m_{\rm s}
    I^{\Lambda}_1(m_{\rm u})I^{\Lambda}_1(m_{\rm s}) \label{twoloopcorrect1},\\
    \overline{m}^0_{\rm s}&=&m^0_{\rm s}- 32 K m_{\rm u}^2
    I^{\Lambda}_1(m_{\rm u})^2\label{twoloopcorrect2},
\ea
introduced here to avoid double counting of the
't Hooft interaction in gap equations
(see \cite{EPJA2000,Kleva_92}).
Here $m_{\rm u}$ and $m_{\rm s}$ are constituent quark masses,
and $I_1^{\Lambda}(m_a)$ stands for a regularized integral
over the momentum space. It is convenient to define
all integrals that will appear further in the paper
via the functional ${\cal J}$:
\beq
\JJ{l}{n}{f}=-i\frac{N_c}{(2\pi)^4}\int\!d^4k
\frac{f(\vec{k})\theta (\Lambda^2-\vec{k}^2)}{(m_{\rm u}^2-k^2)^l(m_{\rm s}^2-k^2)^n},
\label{DefI}
\eeq
where $f$ is a product of form factors, and
$N_c=3$ is the number of colors.
Since the integral is
divergent for some values of $l$ and $n$, it is regularized
by a 3-dimensional  cutoff $\Lambda$. Thus the integrals $I_1^{\Lambda}(m_a)$
($a=$u,s) can be defined as follows:
$I_1^{\Lambda}(m_{\rm u})=\JJ{1}{0}{1}$, and
$I_1^{\Lambda}(m_{\rm s})=\JJ{0}{1}{1}$.


After  bosonization of  Lagrangian (\ref{LGus})
we obtain:
\ba
&& \tilde{\mathcal L}(\bar\sigma,\phi)=
\tilde L_{\rm G}(\bar\sigma,\phi)-i \Tr \ln \biggl\{ i{\hat \partial}-\bar m^0\nonumber\\
&&\quad+\sum_{i=1}^2\sum_{a=1}^9\!\tau_a g_{a,i}
(\bar\sigma_{a,i} + i\gamma_5 \sqrt{Z}\phi_{a,i})f_i^a \biggr\},
\ea
where
\ba
&& \tilde L_{\rm G}(\bar\sigma,\phi)=\nonumber\\
&&\quad=-\frac12\sum_{a,b=1}^9
g_{a,1}\bar\sigma_{a,1}\left(G^{(-)}\right)^{-1}_{ab}
g_{b,1}\bar\sigma_{b,1}\nonumber\\
&&\quad-\frac{Z}{2}\sum_{a,b=1}^9
g_{a,1}\phi_{a,1}\left(G^{(+)}\right)^{-1}_{ab}
g_{b,1}\phi_{b,1}\nonumber\\
&&\quad-\frac{1}{2G}
\sum_{a=1}^9g_{a,2}^2(\sigma_{a,2}^2+\phi_{a,2}^2).
\ea
As it follows from our further calculations of
quark loop diagrams,
the vacuum expectation values (VEV)
of the fields $\bar\sigma_{8,1}$ and $\bar\sigma_{9,1}$
are not equal to zero, while $\vev{\bar\sigma_{a,1}}=0$, $(a=1,\ldots,7)$.
Therefore, it is necessary to introduce new fields
$\sigma_{a,i}$ with  zero VEV
$\vev{\sigma_{8,i}}=\vev{\sigma_{9,i}}=0$,
using  the following relations:
\ba
g_{8,1}\bar\sigma_{8,1}-\mbar^0_{\rm u}&=&
    g_{8,1}\sigma_{8,1}-m_{\rm u},\nonumber\\
g_{9,1}\bar\sigma_{9,1}+\frac{\mbar^0_{\rm s}}{\sqrt{2}}&=&
    g_{9,1}\sigma_{9,1}+\frac{m_{\rm s}}{\sqrt{2}}.
    \label{sigmashift}
\ea
This is connected with the existence of tadpole diagrams for
the ground meson states,
VEV taken from (\ref{sigmashift}) give gap equations
connecting current and constituent quark masses
(see (\ref{gap:eq:u}) and (\ref{gap:eq:s}) in sect.~4).
This is a consequence of spontaneous breaking of chiral symmetry
(SBCS).
As a result  (see, \textit{e.g.}, \cite{EPJA2000,Cimen_99}),
we obtain:
\ba
&& {\mathcal L}(\sigma,\phi)=L_{\rm G}(\sigma,\phi)\nonumber\\
&&-i~{\rm Tr}\ln \left\{ i{\hat \partial}\!-\!m\!+\!\!
\sum_{i=1}^2\sum_{a=1}^9\!\tau_a g_{a,i}(\sigma_{a,i}\! + i\gamma_5\sqrt{Z} \phi_{a,i})f_i^a \right\}\!%
=\nonumber\\
&&=L_{\rm kin}(\sigma,\phi)+L_{\rm G}(\sigma,\phi)+L_{\rm loop}(\sigma,\phi).
\label{Lagr:bosonized:r}
\ea
The term $ L_{\rm G}(\sigma,\phi)$ is
\ba
&& L_{\rm G}(\sigma,\phi)=\nonumber\\
&&\quad=-\frac12\sum_{a,b=1}^9\!
(g_{a,i}\sigma_{a,1}\!-\!\mu_a\!+\!\bar\mu_a^0)\left(G^{(-)}\right)^{-1}_{ab}\nonumber\\
&&\quad\times(g_{b,1}\sigma_{b,1}\!-\!\mu_b\!+\!\bar\mu_b^0)\nonumber\\
&&\quad-
\frac{Z}{2}\sum_{a,b=1}^9
g_{a,1}\phi_{a,1}\left(G^{(+)}\right)^{-1}_{ab}
g_{b,1}\phi_{b,1}\nonumber\\
&&\quad-\frac{1}{2G}\sum_{a=1}^9g_{a,2}^2(\sigma_{a,2}^2+\phi_{a,2}^2).
\label{LG1}
\ea
Here we introduced, for convenience, the constants $\mu_a$ and
$\bar\mu_a^0$ defined as follows: $\mu_a=0, \quad (a=1,\dots ,7)$,
 $\mu_8=m_{\rm u}$, $\mu_9=-m_{\rm s}/\sqrt{2}$ and
$\bar\mu_a^0=0, \quad (a=1,\dots ,7)$,
 $\bar\mu^0_8=\mbar_{\rm u}^0$, $\bar \mu^0_9=-\mbar_{\rm s}^0/\sqrt{2}$.

\begin{figure}[H]
\section*{Figures}
\includegraphics[scale=0.8]{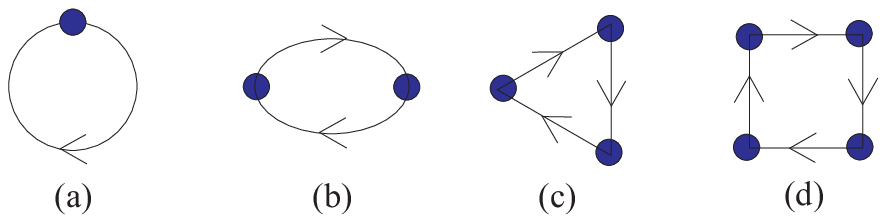}
\caption{}\label{loops}
\end{figure}

The term $L_{\rm kin}(\sigma,\phi)$
contains the kinetic terms and, in the momentum space,
has the following form:
\ba
&&L_{\rm kin}(\sigma,\phi)=\frac{P^2}{2}\sum_{i,j=1}^2\sum_{a=1}^9
\Bigl(\sigma_{a,i}\Gamma_{{\rm S},i j}^a
\sigma_{a,j}\nonumber\\
&&\quad+\phi_{a,i}\Gamma_{{\rm P},i j}^a \phi_{a,j}\Bigr),
\ea
where
\ba
&&\Gamma_{{\rm S(P)},1 1}^a=\Gamma_{{\rm S(P)},2 2}^a=1\nonumber\\
&&\Gamma_{{\rm S(P)},1 2}^a=\Gamma_{{\rm S(P)},2 1}^a=
\gamma_{{\rm S(P)}}^a,
\ea
\beq
\gamma_{{\rm S}}^a=
\left\{\begin{array}{ll}
\frac{\JJ{2}{0}{f_2^a}}{\sqrt{\JJ{2}{0}{1}
\JJ{2}{0}{f_2^a f_2^a}}}&\quad (a=1,2,3,8),\\
\frac{\JJ{1}{1}{f_2^a}}{\sqrt{\JJ{1}{1}{1}
\JJ{1}{1}{f_2^a f_2^a }}}&\quad (a=4,5,6,7),\\
\frac{\JJ{0}{2}{f_2^a}}{\sqrt{\JJ{0}{2}{1}
\JJ{0}{2}{f_2^a f_2^a}}}&\quad (a=9),
\end{array}
\right.
\eeq
\beq
\gamma_{{\rm P}}^a=\gamma_{{\rm S}}^a \sqrt{Z}.
\eeq
The term $L_{\rm loop}(\sigma,\phi)$ is a sum of
one-loop (see Fig.\ref{loops}) quark contributions\footnote{
Here we keep only the terms of an order not higher than 4
(corresponding diagrams are shown in Fig.~\ref{loops}).
}, from which the kinetic term was subtracted:
\ba
&&\!\!L_{\rm loop}(\sigma,\phi)\!=
L_{\rm loop}^{(1)}(\sigma)+
L_{\rm loop}^{(2)}(\sigma,\phi)\nonumber\\
&&+L_{\rm loop}^{(3)}(\sigma,\phi)+
L_{\rm loop}^{(4)}(\sigma,\phi),
\ea
where the superscript in brackets stands for the degree
of fields. Thus, $L_{\rm loop}^{(1)}$ (Fig.~\ref{loops}(a)) contains
the terms linear in the field $\sigma$; $L_{\rm loop}^{(2)}$
(Fig.~\ref{loops}(b)), the quadratic ones, and so on.
For example,
\ba
&&L_{\rm loop}^{(1)}(\sigma,\phi)= 8 m_{\rm u} g_{8,1}I_1^\Lambda(m_{\rm u})\sigma_{8,1}\nonumber\\
&&\quad-4\sqrt{2}m_{\rm s} g_{9,1}I_1^\Lambda(m_{\rm s})\sigma_{9,1},
\ea
\ba
&&L_{\rm loop}^{(2)}(\sigma,\phi)=
    4\sum_{a=1}^3 g_{a,1}^2 I_1^{\Lambda}(m_{\rm u})(\sigma_{a,1}^2+Z\phi_{a,1}^2)\nonumber\\
&&\quad+2\sum_{a=4}^7 g_{a,1}^2 (I_1^{\Lambda}(m_{\rm u})+I_1^{\Lambda}(m_{\rm s}))(\sigma_{a,1}^2+Z\phi_{a,1}^2)\nonumber\\
&&\quad+4g_{8,1}^2 I_1^{\Lambda}(m_{\rm u})(\sigma_{8,1}^2+Z\phi_{8,1}^2)\nonumber\\
&&\quad+4g_{9,1}^2I_1^{\Lambda}(m_{\rm s})(\sigma_{9,1}^2+Z\phi_{9,1}^2)\nonumber\\
&&\quad+4\sum_{a=1}^3 g_{a,2}^2 \JJ{1}{0}{f_2^a f_2^a}(\sigma_{a,2}^2+\phi_{a,2}^2)\nonumber\\
&&\quad+2\sum_{a=4}^7 g_{a,2}^2 (\JJ{1}{0}{f_2^a f_2^a}+\JJ{0}{1}{f_2^a f_2^a})(\sigma_{a,2}^2+\phi_{a,2}^2)\nonumber\\
&&\quad+4g_{8,2}^2\JJ{1}{0}{f_2^8f_2^8}(\sigma_{8,2}^2+\phi_{8,2}^2)\nonumber\\
&&\quad+4g_{9,2}^2\JJ{0}{1}{f_2^9f_2^9}(\sigma_{9,2}^2+\phi_{9,2}^2)\nonumber\\
&&\quad-2\sum_{i,j=1}^2\biggl[
	m_{\rm u}^2\sum_{a=1}^3\sigma_{a,i}\Gamma_{\mathrm S,ij}^a\sigma_{a,j}\nonumber\\
&&\quad+\left(\frac{m_{\rm u}+m_{\rm s}}{2}\right)^2\sum_{a=4}^7\sigma_{a,i}\Gamma_{\mathrm S,ij}^a\sigma_{a,j}\nonumber\\
&&\quad+m_{\rm u}^2\sigma_{8,i}\Gamma_{\mathrm S,ij}^8\sigma_{8,j}+
	m_{\rm s}^2\sigma_{9,i}\Gamma_{\mathrm S,ij}^9\sigma_{9,j}
\biggr].
\ea
The total expressions for $L_{\rm loop}^{(3)}$
and $L_{\rm loop}^{(4)}$ are too lengthy, therefore,
we do not show them here. Instead we will extract parts
from them when they are needed
(see \textit{e.g.\/} Appendix A).

The Yukawa coupling constants $g_{a,i}$ describing  the
interaction of quarks and mesons appear as a result of
renormalization of meson fields
(see \cite{IJMPA1999,YF,ECHAJA2000,VolkovWeiss,Volk1997,Volk_86}
for details):
\ba
&& g_{a,1}^2=[4\JJ{2}{0}{1}]^{-1}, \quad (a=1,2,3,8),\nonumber\\
&&g_{a,1}^2=[4\JJ{1}{1}{1}]^{-1}, \quad (a=4,5,6,7),\nonumber \\
&&g_{9,1}^2=[4\JJ{0}{2}{1}]^{-1}.
\label{ga_0}
\ea
\ba
&& g_{a,2}^2=[4\JJ{2}{0}{f_2^a f_2^a}]^{-1}, \quad (a=1,2,3,8),\nonumber\\
&&g_{a,2}^2=[4\JJ{1}{1}{f_2^af_2^a}]^{-1}, \quad (a=4,5,6,7),\nonumber \\
&&g_{9,2}^2=[4\JJ{0}{2}{f_2^9 f_2^9}]^{-1}.
\label{ga_0ex}
\ea
For the pseudoscalar meson fields,
$\pi$-$A_1$-transitions
lead to the  factor $Z$, describing
an additional renormalization of pseudoscalar
meson fields, with $M_{A_1}$ being
the mass of the axial-vector meson (see \cite{Volk1997,Volk_86}):
\beq
Z=\left(1-\frac{6m_{\rm u}}{M_{A_1}^2}\right)^{-1}\approx 1.46 \label{Z} .
\eeq
For the radially excited pseudoscalar states a similar
renormalization also takes place, but in this case
the renormalization factor turns out to be approximately
equal to unit, so it is omitted in our calculations
(see \cite{Volk1997}).

\section{Introducing the dilaton}

According to the prescription described
in \cite{EPJA2000,YF2000},
we introduce the dilaton field into Lagrangian (\ref{Lagr:bosonized:r})
as follows: the dimensional model parameters
$G$, $\Lambda$, $m_a$,  and $K$  are replaced by the following rule:
$G\to G(\chi_c/\chi)^2$,
$\Lambda\to\Lambda (\chi/\chi_c)$,
$m_a\to m_a(\chi/\chi_c)$, $K\to K(\chi_c/\chi)^5$,
where $\chi$ is the dilaton field with VEV $\chi_c$.
We also define the field $\chi'$ as
the difference $\chi'=\chi-\chi_c$ that has zero VEV.
Below the effective meson Lagrangian is expanded
in terms of $\chi'$ when calculating the mass terms
and vertices describing the interaction of mesons.

The current quark masses break scale invariance and,
therefore, should not be multiplied by the dilaton field.
The modified current quark masses $\mbar_a^0$ are also
not multiplied by the dilaton field.
Finally, we come to the Lagrangian:
\ba
&&\bar{\mathcal L}(\sigma,\phi,\chi)=
L_{\rm kin}(\sigma,\phi)+
\bar L_{\rm G}(\sigma,\phi,\chi)
+\bar L_{\rm loop}(\sigma,\phi,\chi)\nonumber\\
&&\qquad+{\mathcal L}(\chi)+\Delta L_{\rm an}(\sigma,\phi,\chi).
\label{Lagr:bosonized:chi}
\ea
The term $L_{\rm kin}$ remains unchanged, as it is already
scale-invariant.

Here, the term $\bar L_{\rm G}(\sigma,\phi,\chi)$ is
\ba
&&\bar L_{\rm G}(\sigma,\phi,\chi)=\nonumber\\
&&\quad=-\frac12\left({\chi\over\chi_c}\right)^2\!\!\sum_{a,b=1}^9
\!\left(g_{a,1}\sigma_{a,1}-\mu_a{\chi\over\chi_c}+
\bar\mu^0_a\right)\left(G^{(-)}\right)^{-1}_{ab}\nonumber\\
&&\quad\times
    \left(g_{b,1}\sigma_{b,1}-\mu_b{\chi\over\chi_c}
+\bar\mu^0_b\right)\nonumber\\
&&\quad -\frac{Z}{2}\left({\chi\over\chi_c}\right)^2\sum_{a,b=1}^9
g_{a,1}\phi_{a,1}\left(G^{(+)}\right)^{-1}_{ab}g_{b,1}\phi_{b,1}\nonumber\\
&&\quad -\frac{1}{2G}\left(\frac{\chi}{\chi_c}\right)^2
\sum_{a=1}^{9}g_{a,2}^2(\sigma_{a,2}^2+\phi_{a,2}^2).
\label{LGr}
\ea
Expanding (\ref{LGr})
in a power series of $\chi$, we can extract a term that is of order $\chi^4$.
It can be absorbed by the term in the pure dilaton potential
(see (\ref{chi}) below)
which has the same degree of $\chi$. This does not bring
essential changes, because  such terms are scale-invariant
and therefore do not contribute to the divergence of the
dilatation current.
This would lead only to a redefinition of
the constants  $\chi$ and $B$ of
the potential (\ref{chi}).

The sum of one-loop quark diagrams is denoted as
$\bar L_{\rm loop}$:
\ba
&&\bar L_{\rm loop}(\sigma,\phi,\chi)=
L_{\rm loop}^{(1)}(\sigma)\left(\frac{\chi}{\chi_c}\right)^3+
L_{\rm loop}^{(2)}(\sigma,\phi)\left(\frac{\chi}{\chi_c}\right)^2\nonumber\\
&&\quad+L_{\rm loop}^{(3)}(\sigma,\phi)\frac{\chi}{\chi_c}+
L_{\rm loop}^{(4)}(\sigma,\phi).
\label{qloops}
\ea
Here ${\mathcal L}(\chi)$ is the pure dilaton Lagrangian
\beq
{\mathcal L}(\chi)=\frac{P^2}{2}\chi^2-V(\chi)
\eeq
with the  potential
\ba
V({\chi})=B\left({\chi\over {\chi}_0} \right)^4\left[ \ln \left({\chi\over
{\chi}_0} \right)^4 -1 \right] \label{chi}
\ea
that has a
minimum at $\chi = \chi_0$, and the parameter $B$
represents the vacuum
energy when there are no quarks. The kinetic term is given
in the momentum space, $P$ being the momentum of the dilaton.

Note that Lagrangian (\ref{Lagr:bosonized:r}) implicitly
contains the term $L_{\rm an}$ that is
induced by gluon anomalies:
\beq
L_{\rm an}(\bar\sigma,\phi)=
-\hps\phi_0^2+\hsc\bar\sigma_0^2, \label{anomalynotscaled}
\eeq
where $\phi_0$ and $\bar\sigma_0\quad
(\vev{\sigma_0} \not =0)$
are pseudoscalar and scalar meson isosinglets, respectively;
and $\hps, \hsc $ are constants;
$\phi_0=\sqrt{2/3}\phi_{8,1}-\sqrt{1/3}\phi_{9,1}$,
$\bar\sigma_0=\sqrt{2/3}\bar\sigma_{8,1}-\sqrt{1/3}\bar\sigma_{9,1}$, where
$\phi_{8,1}$ and $\bar\sigma_{8,1}$
($\vev{\bar\sigma_{8,1}}\not=0$) consist of $u$-quarks;
and $\phi_{9,1}$, $\bar\sigma_{9,1}$
($\vev{\bar\sigma_{9,1}}\not=0$), of $s$-quarks.
In our model, the 't~Hooft interaction is responsible for
the appearance of these terms.

When  the procedure of the scale invariance restoration is applied
to  Lagrangian (\ref{Lagr:bosonized:r}), the term $L_{\rm an}$
also becomes scale-invariant. To avoid this, one should
subtract this part in the scale-invariant form and add
it in a scale-breaking (SB) form. This is achieved by
including the  term $\Delta L_{\rm an}$:
\beq
\Delta L_{\rm an}(\sigma,\phi,\chi)=
-L_{\rm an}(\bar\sigma,\phi)\left(\frac{\chi}{\chi_c}\right)^2
+L_{\rm an}^{\rm SB}(\sigma,\phi,\chi). \label{Lan}
\eeq
Let us define the scale-breaking term $L_{\rm an}^{\rm SB}$.
The coefficients $\hsc$ and $\hps$ in (\ref{anomalynotscaled})
can be determined by
comparing them with the  terms in (\ref{LG1})
that describe the singlet-octet mixing\footnote{
The singlet-octet mixing is fully determined by the 't Hooft
interaction.
}.
We obtain
\ba
\hps&=&-\frac{3}{2\sqrt{2}} g_{8,1} g_{9,1} Z \left(G^{(+)}\right)^{-1}_{89},\\
\hsc&=&\frac{3}{2\sqrt{2}} g_{8,1} g_{9,1}  \left(G^{(-)}\right)^{-1}_{89}.
\ea
If  these terms were to be made scale-invariant,
one should insert $(\chi/\chi_c)^2$ into them (see (\ref{Lan})).
However, as the gluon anomalies break scale invariance, we
introduce the dilaton field into these terms in a more complicated way.
The inverse matrix  elements $\left(G^{(+)}\right)^{-1}_{ab}$ and
$\left(G^{(-)}\right)^{-1}_{ab}$,
\beq
\left(G^{(+)}\right)^{-1}_{89}={-4\sqrt{2}m_{\rm u} K I_1^\Lambda(m_{\rm u})\over
G^{(+)}_{88}G^{(+)}_{99}-\left(G^{(+)}_{89}\right)^2},
\eeq
\beq
\left(G^{(-)}\right)^{-1}_{89}={4\sqrt{2}m_{\rm u} K I_1^\Lambda(m_{\rm u})\over
G^{(-)}_{88}G^{(-)}_{99}-\left(G^{(-)}_{89}\right)^2},
\eeq
are determined by two different interactions.
The numerators are fully defined by the 't Hooft
interaction that leads to anomalous terms (\ref{anomalynotscaled})
breaking scale invariance, therefore, we do not introduce
here dilaton fields. The denominators are determined by the
constant $G$ describing the standard NJL four-quark interaction,
and the dilaton field is inserted into it,
according to the prescription given above.
Finally, we come to the following form of $L_{\rm an}^{\rm SB}$:
\ba
&&\!\!\!\!L_{\rm an}^{\rm SB}(\sigma,\phi,\chi)\!\!=\!\!\!\left[\!\!-\hps\phi_0^{2}\!+\!
\hsc\!\!\left(\!\sigma_0\!-\!\!F_0\!\frac{\chi}{\chi_c}\!+\!F_0^0\!\right)^{\!\!2}\right]
\!\!\!\left(\!\frac{\chi}{\chi_c}\!\right)^{\!\!4}\!\!\!,\\
&& F_0=
\frac{\sqrt{2}m_{\rm u}}{\sqrt{3}g_{8,1}}+\!\frac{m_{\rm s}}{\sqrt{6}g_{9,1}},\quad
F_0^0=\frac{\sqrt{2}\mbar_{\rm u}^0}{\sqrt{3}g_{8,1}}%
+\!\frac{\mbar_{\rm s}^0}{\sqrt{6}g_{9,1}}.
\ea
From it, we immediately obtain the term $\Delta L_{\rm an}$:
\ba
&&\Delta L_{\rm an}
=\left[\hps\phi_0^{2}-\hsc
\left(\sigma_0-F_0\frac{\chi}{\chi_c}+F_0^0
\right)^2\right]\nonumber\\
&&\quad\times\left(
\frac{\chi}{\chi_c}\right)^2\left[1-\left(\frac{\chi}{\chi_c}\right)^2\right]
\label{Deltaanomalynotscaled}.
\ea

\section{Equations}

Let us now consider VEV of the
divergence of the dilatation current $S^\mu$ \cite{Kusaka,EPJA2000}
calculated from the potential of Lagrangian (\ref{Lagr:bosonized:chi}):
\ba
 &&\langle\partial_{\mu}S^{\mu}\rangle\!=\!\left[\sum_{i=1}^2
 \sum_{a=1}^9\left(
\!\sigma_{a,i}\!{\partial V\over\partial \sigma_{a,i}}\!+\!
\phi_{a,i}\!{\partial V\over\partial \phi_{a,i}}\right)\!\right.\nonumber\\
&&\quad+\!\chi\!{\partial V\over\partial \chi}\!-\!4V\Biggr]
\!\Biggr\vert_{\begin{array}{l}\scriptstyle \chi=\chi_c\, \\[-1.5mm]
\scriptstyle\sigma=0 \\[-1.5mm] \scriptstyle\phi=0
\end{array}}\kern-3mm=\nonumber\\
&&\quad=4B\left({\chi_c\over\chi_0}\right)^4\!\!\!\!%
-\!2\hsc\!\left(F_0-F_0^0\right)^2\!
-\kern-3mm\sum_{q=\rm u,d,s}\!\!\!\mbar^0_q\vev{qq}.
\label{dilaton:current}
\ea

Here $V=V(\chi)+\bar V(\sigma,\phi,\chi)$,
and $\bar V(\sigma,\phi,\chi)$
is the potential part of Lagrangian
$\bar{\mathcal L}(\sigma,\phi,\chi)$ (see (\ref{Lagr:bosonized:chi}))
that does not contain the pure dilaton potential (\ref{chi}).
In the expression given in  (\ref{dilaton:current}),
the following relation  of the quark condensates
to integrals $I_1^\Lambda(m_{\rm u})$ and $I_1^\Lambda(m_{\rm s})$
was used:
\beq
4m_q I^\Lambda_1(m_q)=-\vev{\bar qq},
\qquad (q=\rm u,d,s), \label{I1toQQ}
\eeq
and that these integrals are connected with constants $G^{(-)}_{ab}$
through gap equations, as it will be shown below
(see (\ref{gapeqbegin}) and (\ref{gapeq2})).
Comparing (\ref{dilaton:current}) with the QCD expression
\beq
\langle\partial_{\mu}S^{\mu}\rangle=
\Cg-\sum_{q={\rm u,d,s}}m^0_q\vev{\bar qq}, \label{Ward}
\eeq
where
\beq
\Cg=\left(\frac{11 N_c}{24}-\frac{N_{\rm f}}{12}\right)
\vev{\frac{\alpha}{\pi}\left(G_{\mu\nu}^a\right)^2},
\eeq
and  $N_{\rm f}$ is the number
of flavours, $\vev{\frac{\alpha}{\pi}(G_{\mu\nu}^a)^2}$
and $\vev{\bar qq}$ are the gluon and quark condensates with
$\alpha$ being the QCD constant of strong interaction,
one can see that the term $\sum m^0_q\langle\bar qq\rangle$
on the right-hand side of (\ref{Ward}) is canceled by
the corresponding contribution from current quark masses
on the right-hand side of (\ref{dilaton:current}).
Equating the right-hand sides of (\ref{dilaton:current}) and (\ref{Ward}),
\ba
&&\Cg-\!\sum_{q=\rm u,d,s}\!m^0_q\vev{\bar qq}=\nonumber\\
&&\quad=4B\left({\chi_c\over\chi_0}\right)^4\!\!\!
-2\hsc\left(F_0-F_0^0\right)^2\!\!\!-\kern-3mm\sum_{q=\rm u,d,s}\!\!\mbar^0_q\vev{\bar qq},
\ea
we obtain the correspondence
\ba
&&\Cg= 4B\left(\chi_c\over\chi_0\right)^4\!+
\!\!\sum_{a,b=8}^9(\bar\mu_a^0-\mu_a^0)\left(G^{(-)}\right)_{ab}^{-1}(\mu_b-\bar\mu_b^0)\nonumber\\
&&\quad-2\hsc\left(F_0-F_0^0\right)^2, \label{Cg}
\ea
where $\mu_a^0=0\quad (a=1,\dots 7)$, $\mu_8^0=m^0_{\rm u}$,
and $\mu_9^0=-m_{\rm s}^0/\sqrt{2}$.
This equation relates the gluon condensate, whose value is taken
from other sources (see, \textit{e.g.}, \cite{Narison96}),
to the model parameter $B$. The next step is to investigate
gap equations.

As usual,
gap equations follow from the requirement that
the terms linear in $\sigma$ and $\chi'$
should be absent in the effective Lagrangian:
\ba
&&{{\delta}\bar{\mathcal L}\over {\delta}\sigma_{8,1}}
\biggr\vert_{\minimumpoint}\!\!\!=
{{\delta}\bar{\mathcal L}\over {\delta}\sigma_{9,1}}
\biggr\vert_{\minimumpoint}\!\!\!=
{{\delta}\bar{\mathcal L}\over {\delta}\chi}
\biggr\vert_{\minimumpoint}\!\!\!=\nonumber\\
&&={{\delta}\bar{\mathcal L}\over {\delta}\sigma_{8,2}}
\biggr\vert_{\minimumpoint}\!\!\!=
{{\delta}\bar{\mathcal L}\over {\delta}\sigma_{9,2}}
\biggr\vert_{\minimumpoint}=0.
\label{var}
\ea
For the ground states  of quarkonia ($\sigma_{a,1}$)
and the dilaton field $\chi'$,
this leads to the following equations:
\ba
(m_{\rm u}-\bar{m}_{\rm u}^0)\left(G^{(-)}\right)^{-1}_{88} - {{m_{\rm u}-\bar{m}_{\rm u}^0}\over \sqrt2}\left(G^{(-)}\right)^{-1}_{89}\quad&&\nonumber\\
\quad-8m_{\rm u}I^\Lambda_1(m_{\rm u})= 0, && \label{gapeqbegin} \\
(m_{\rm s}-\bar{m}_{\rm s}^0)\left(G^{(-)}\right)^{-1}_{99} - {\sqrt2}(m_{\rm s}-\bar{m}_{\rm s}^0)\left(G^{(-)}\right)^{-1}_{98}&&\quad\nonumber\\
-8 m_{\rm s}I^\Lambda_1(m_{\rm s}) = 0, &&\label{gapeq2} \\
4B\left({\chi_c\over {\chi}_0} \right)^3{1\over \chi_0}
\ln \left({\chi_c\over {\chi}_0} \right)^4 \hspace{3cm} &&\nonumber\\
+{1\over \chi_c}\sum_{a,b=8}^9
\bar\mu_a^0\left(G^{(-)}\right)_{ab}^{-1}(\bar\mu_b^0-3\mu_b)-\quad&&\nonumber\\
-\frac{2\hsc}{\chi_c}\left(F_0-F_0^0\right)^2= 0. &&
\label{Gapeqs}
\ea
Using
(\ref{twoloopcorrect1}) and (\ref{twoloopcorrect2}), one can rewrite
equations (\ref{gapeqbegin}) and (\ref{gapeq2})  in
the well-known form \cite{Kleva_92}:
\ba
m_{\rm u}^0&=&m_{\rm u}-8 G m_{\rm u} I_1^{\Lambda}(m_{\rm u})\nonumber\\
&&-32K m_{\rm u} m_{\rm s} I_1^\Lambda(m_{\rm u})I_1^\Lambda(m_{\rm s}),
\label{gap:eq:u}\\
m_{\rm s}^0&=&m_{\rm s}-8 G m_{\rm s} I_1^{\Lambda}(m_{\rm s})-32K(m_{\rm u} I_1^\Lambda(m_{\rm u}))^2.
\label{gap:eq:s}
\ea

For the excited states ($\sigma_{a,2}$), we require
that the corresponding gap equations have the trivial
solution, \textit{i.e.},
$\sigma_{a,2}$ do not acquire VEV.
This is one of the possible
particular solutions of equations (\ref{var}).
An advantage of such a solution is that in this case
the quark condensates and constituent quark masses
remain unchanged after introducing radially excited states.
This solution surely exists if the tadpole diagram
(Fig.~\ref{loops}(a)) for
the excited scalar is equal to zero
(see \cite{ECHAJA2000,VolkovWeiss}).
This leads to the condition:
\beq
\JJ{1}{0}{f_2^a}=\JJ{0}{1}{f_2^a}=0
\label{zerotadpole}.
\eeq
The calculation of the second variation of
the effective potential will ensure us that
the solution that we have chosen give the
minimum of the potential.

The integrals in (\ref{zerotadpole}) depend on $\Lambda$,
$m_{\rm u}$, and $m_{\rm s}$. The form factors in them depend on
the external  $c_a$ and slope $d_a$ parameters. The external
parameter factors out, and the only
possibility to satisfy (\ref{zerotadpole}) is to chose
appropriate values of $d_a$. Insofar as there two different
conditions (\ref{zerotadpole}), we obtain two different
magnitudes: $d_{\rm u}$, $d_{\rm s}$.
The difference appears from the difference
between the constituent masses of $u$ and $s$ quarks.

To determine the masses of quarkonia and of the glueball, let
us consider the part of Lagrangian (\ref{Lagr:bosonized:chi})
which is quadratic in fields $\sigma$ and $\chi'$
and which is denoted by $L^{(2)}$:
\ba
&&L^{(2)}(\sigma,\phi,\chi')=\frac{1}{2}
\sum_{i,j=1}^2\biggl[\sum_{a=1}^3(P^2-4m_{\rm u}^2)
\sigma_{a,i}\Gamma_{{\rm S},i j}^a
\sigma_{a,j}\nonumber\\
&&\quad +\sum_{a=4}^7(P^2-(m_{\rm u}+m_{\rm s})^2)
\sigma_{a,i}\Gamma_{{\rm S},i j}^a
\sigma_{a,j}\nonumber\\
&&\quad +(P^2-4m_{\rm u}^2)
\sigma_{8,i}\Gamma_{{\rm S},i j}^8
\sigma_{8,j}\!+\!(P^2-4m_{\rm s}^2)
\sigma_{9,i}\Gamma_{{\rm S},i j}^9
\sigma_{9,j}\biggr]\nonumber\\
&&\quad -{1\over 2}g_{8,1}^2\left[\left(G^{(-)}\right)^{-1}_{88}
-8I^\Lambda_1(m_{\rm u})\right]{\sigma}^2_{8,1} \nonumber \\
&&\quad-{1\over 2}g_{9,1}^2\left[\left(G^{(-)}\right)^{-1}_{99} -8I^\Lambda_1(m_{\rm s})\right]{\sigma}^2_{9,1}
  \nonumber \\
&&\quad-{1\over 2}g_{8,2}^2\left[1/G
-8\JJ{1}{0}{f_2^8f_2^8}\right]{\sigma}^2_{8,2} \nonumber \\
&&\quad-{1\over 2}g_{9,2}^2\left[1/G -8\JJ{0}{1}{f_2^9f_2^9}\right]{\sigma}^2_{9,2}
  \nonumber \\
&&\quad-g_{8,1}g_{9,1}\left(G^{(-)}\right)^{-1}_{89}{\sigma}_{8,1}{\sigma}_{9,1}
-\frac{M_{\rm g}^2 {\chi'}^2}{2}\nonumber\\
&&\quad+\!\!\!\sum_{a,b=8}^9\frac{\bar\mu_a^0}{\chi_c}(G^{(-)})^{-1}_{ab}
g_{b,1}\sigma_{b,1}\chi'\nonumber\\
&&\quad+\frac{4\hsc (F_0-F_0^0)}{\chi_c\sqrt{3}}
\left({\sigma_{9,1}}-\sigma_{8,1}\sqrt{2}\right)\chi'
\label{BAchi},
\ea
where
\ba
&&M_{\rm g}^2=\frac{1}{\chi_c^2}\biggl(4\Cg+\sum_{a,b=8}^{9}
\bar\mu^0_a \left(G^{(-)}\right)^{-1}_{ab}(2\bar\mu^0_b-\mu_b)\nonumber\\
&&\qquad+\sum_{a,b=8}^{9}4\mu^0_a \left(G^{(-)}\right)^{-1}_{ab}(\mu_b-\bar\mu_b^0)\nonumber\\
&&\qquad-4\hsc F_0^2+ 4\hsc (F_0^0)^2\biggr)
\ea
is the glueball mass before taking account of mixing effects.
Here, the gap equations and equation (\ref{Cg}) are
taken into account.

From this Lagrangian, after diagonalization, we obtain
the masses of five scalar isoscalar meson states:
$\sigmaI$, $\sigmaII$, $\sigmaIII$, $\sigmaIV$,
and $\sigmaV$ and a matrix of mixing coefficients $b$ that
connects the nondiagonalized fields $\sigma_{8,1},
\sigma_{9,1}, \sigma_{8,2}, \sigma_{9,2}, \chi'$
with the physical ones $\sigmaI,
\sigmaII,\sigmaIII, \sigmaIV, \sigmaV$:
\beq
\left(\!
\begin{array}{c}
\sigma_{8,1}\\ \sigma_{9,1}\\\sigma_{8,2}\\ \sigma_{9,2} \\ \chi'
\end{array}
\!\right)\kern-1.5mm=\kern-1.5mm
\left(
\begin{array}{lllll}
b_{\sigma_{8,1}\sigmaI} & b_{\sigma_{8,1}\sigmaII}&b_{\sigma_{8,1}\sigmaIII}&b_{\sigma_{8,1}\sigmaIV}&b_{\sigma_{8,1}\sigmaV}\\
b_{\sigma_{9,1}\sigmaI} & b_{\sigma_{9,1}\sigmaII}&b_{\sigma_{9,1}\sigmaIII}&b_{\sigma_{9,1}\sigmaIV}&b_{\sigma_{9,1}\sigmaV}\\
b_{\sigma_{8,2}\sigmaI} & b_{\sigma_{8,2}\sigmaII}&b_{\sigma_{8,2}\sigmaIII}&b_{\sigma_{8,2}\sigmaIV}&b_{\sigma_{8,2}\sigmaV}\\
b_{\sigma_{9,2}\sigmaI} & b_{\sigma_{9,2}\sigmaII}&b_{\sigma_{9,2}\sigmaIII}&b_{\sigma_{9,2}\sigmaIV}&b_{\sigma_{9,2}\sigmaV}\\
b_{\chi'\sigmaI} & b_{\chi'\sigmaII}&b_{\chi'\sigmaIII}& b_{\chi'\sigmaIV}&b_{\chi'\sigmaV}\\
\end{array}%
\!\right)\kern-1.5mm\left(%
\!\begin{array}{c}
\sigmaI\\ \sigmaII \\ \sigmaIII \\ \sigmaIV \\ \sigmaV
\end{array}
\!\!\right)\!.
\label{mixingcoeff}
\eeq
The values of matrix elements are given in Table~\ref{mix}.

\section{Model parameters and numerical estimates}

The basic parameters of our model are $G$,  $\Lambda$, $m_{\rm u}$,
and $m_{\rm s}$. They are fixed by the pion weak decay constant
$F_\pi=93$ MeV, the $\rho$ meson decay constant
$g_\rho\approx 6.14$
describes the decay of a $\rho$-meson into 2 pions,
and the masses of pion and kaon \cite{Volk_86,Kikkawa,VolkovEbert}.
To fix $\Lambda$ and $m_{\rm u}$, the Goldberger-Treiman relation
$g_{\rm u} F_\pi\sqrt{Z}=m_{\rm u}$ and the equation $g_\rho=\sqrt{6}g_{\rm u}$
are used.
The parameter $G$ is determined by the pion mass;
and $m_{\rm s}$, by the kaon mass.
Their values do not change both after the radially excited states
\cite{IJMPA1999,YF,ECHAJA2000,VolkovWeiss,Volk1997}
and  the dilaton fields are introduced \cite{EPJA2000,YF2000}:
\ba
&&m_{\rm u}=280\;\mbox{MeV},\;m_{\rm s}=417\;\mbox{MeV},\;\Lambda =1.03\;\mbox{
GeV},\nonumber\\
&&G=3.2017\;\mbox{GeV}^{-2}. \label{paramet}
\ea

To have a correct description  of  $\eta$ and $\eta'$,
one should fix the 't Hooft interaction constant by
the masses of $\eta$  and $\eta'$.
The lower bound for the lightest
scalar meson mass is
 also  taken into account here.
As a result, for the model masses we obtain:
$M_\eta\approx 500$ MeV, $M_{\eta'}\approx 870$ MeV, and
for $K$:
\beq
K=4.4\; \mbox{GeV}^{-5}.
\eeq

After  introducing the radially excited states
into the isoscalar sector,
there appear four form factor parameters
$c_{8}, c_{9}$ and $d_{\rm u} (\equiv d_8) , d_{\rm s} (\equiv d_9)$\footnote{
To calculate the widths of the decays of scalars into
pseudoscalars, one needs an additional slope parameter
$d_{\rm us}$ and external parameters for the pion and kaon
$c_\pi, c_K$, whose values are given in  Appendix A.
}.
The slope parameters $d_{\rm u}$ and $d_{\rm s}$  are fixed
by the requirement that the tadpole diagrams related to
the excited states must be equal to zero (\ref{zerotadpole}).
As a result, we obtain:
\beq
d_{\rm u}= -1.77\;\mbox{GeV}^{-2};\quad d_{\rm s}= -1.72\;\mbox{GeV}^{-2}.
\eeq
The external form factor parameters $c_{8}$ and $c_{9}$ are free and
are fixed by masses of
radially excited pseudoscalar mesons $\eta(1295)$ and $\eta(1440)$:
\beq
c_{8}=1.45,\; c_{9}=1.59.
\label{c8c9}
\eeq
Due to the chiral symmetry of Lagrangian (\ref{njl}),
the same values of the form factor
parameters are used for the scalar mesons.

After the dilaton is introduced, new three parameters
$\chi_0$, $\chi_c$, and $B$ appear. To fix the new parameters,
one should use equations (\ref{Cg}), (\ref{Gapeqs}), and
the physical glueball mass.
As a result, we  obtain for $\chi_0$ and $B$:
\ba
&&\chi_0=\chi_c \exp \Biggl\{-\biggl[\!
\sum_{a,b=8}^{9}\!
\bar\mu^0_a \left(G^{(-)}\right)^{-1}_{ab}(3\mu_b-\bar\mu^0_b)+
\nonumber\\
&&\quad +2\hsc\left(F_0\!-\!F_0^0\right)^2\biggr]\nonumber\\
&&\quad\Bigr/4\biggl[\Cg
-\sum_{a,b=8}^9\!(\bar\mu^0_a\!-\!\mu^0_a)\!\left(G^{(-)}\right)^{-1}_{ab}\!(\mu_b\!-\!\bar\mu^0_b)\!+\nonumber\\
&&\quad+2\hsc\left(F_0-F_0^0\right)^2\biggr]\Biggr\},
\ea
\ba
&&B=\frac14\biggl(\Cg\!-\!\sum_{a,b=8}^9\!(\bar\mu^0_a\!-\!\mu^0_a)\!\left(G^{(-)}\right)^{-1}_{ab}
\!(\mu_b\!-\!\bar\mu^0_b)+\nonumber\\
&&\quad +2\hsc\left(F_0-F_0^0\right)^2\biggr)
\left(\frac{\chi_0}{\chi_c}\right)^4.
\ea

We adjust the parameter $\chi_c$, so that
the mass of the  scalar meson state $\sigmaIV$ would be
close to 1500 MeV ($\chi_c=0.219$ GeV)\footnote{
To reach more close agreement with experimental data
in the description of strong decays of $\sigmaIV$,
we chose the model value of $M_{\sigmaIV}=1550$ MeV
(mass + half-width)
}. For the constants $\chi_0$ and $B$ we have: $\chi_0=203$ MeV,
$B=0.007$ GeV$^4$.
We found that, if the state
$f_0(1710)$ is supposed to be the glueball,
the result turns out to be in worse
agreement with experiment.
The masses of scalar isoscalar mesons calculated
in our model together with their experimental values
are given in Table~\ref{T:spectr}.

\section{Strong decays of scalar isoscalar mesons}

Once all parameters are fixed, we can  estimate
the decay widths for the main modes of  strong decays
of scalar isoscalar mesons: $\sigma_l\to\pi\pi$,$KK$,
$\eta\eta$, $\eta\eta'$, and  $4\pi$ ($2\sigma, \sigma2\pi\to 4\pi$),
where $l=\rm I, II, III, IV$, and V.

Note  that, in the energy region under consideration (up to 1.7 GeV),
we work on the brim of the validity of exploiting the chiral symmetry
and scale invariance that were used to construct our effective Lagrangian.
Thus, our results should be considered rather as qualitative.

Let us start with the decay of a scalar isoscalar
meson into a pair of pions. The corresponding amplitude has
the  form:
\beq
A_{\sigma_{l}\to\pi\pi}=
    A_{\sigma_{l}\to\pi\pi}^{(1)}+
    A_{\sigma_{l}\to\pi\pi}^{(2)},
    \label{amplitudeSigmaTo2pi}
\eeq
where the first part comes from contact terms of Lagrangian
(\ref{Lagr:bosonized:chi}) that describe the decay of
the glueball into pions. These terms come from $\bar L_{G}(\sigma,\phi,\chi)$
and $(\chi/\chi_c)^2L_{\rm loop}^{(2)}(\sigma,\phi)$
(see (\ref{LGr}) and (\ref{qloops})).
They turn into
the pion mass term if $\chi=\chi_c$. Expanding around $\chi=\chi_c$
in terms of $\chi'$ and choosing the term linear in $\chi'$,
we obtain, after the mixing effects are taken into account,
the following:
\beq
A_{\sigma_{l}\to\pi\pi}^{(1)}=-\frac{M_\pi^2}{\chi_c}b_{\chi'\sigma_{l}},
\eeq
where
 $M_\pi$ is the pion mass, and $b_{\chi'\sigma_{l}}$ is a
mixing coefficient (see (\ref{mixingcoeff}) and Table \ref{mix}).
The second contribution $A_{\sigma_{l}\to\pi\pi}^{(2)}$
describes the decay of the quarkonium part of $\sigma_{l}$
and is determined by triangle quark loop diagrams
(see Figs.~\ref{loops}(c) and \ref{trigs}).
For details of their calculation see Appendix A.
As a result, we obtain the following widths for
decays of scalar isoscalar mesons into two pions:
\ba
\Decay{\sigmaI\to\pi\pi}{}&\approx & 600 \mev,\nonumber\\
\Decay{\sigmaII\to\pi\pi}{}&\approx & 36 \mev (20 \mev),\nonumber\\
\Decay{\sigmaIII\to\pi\pi}{}&\approx & 680 \mev (480 \mev),\nonumber\\
\Decay{\sigmaIV\to\pi\pi}{}&\approx & 100 \mev, \nonumber\\
\Decay{\sigmaV\to\pi\pi}{}&\approx & 0.3 \mev,
\ea
To calculate decay widths, we used the model masses of
scalar mesons. For the state
$\sigmaII$ hereafter  we give
in brackets the values obtained for
its experimental mass.
Concerning the state $\sigmaIII$, the values in brackets
correspond to calculations
performed for the lowest experimental
limit for its mass (1200 MeV).  Note that in the last
two cases the widths are noticeably smaller than
those derived for the model masses.

Decays of scalar isoscalar mesons into kaons are described
by the amplitude:
\beq
A_{\sigma_{l}\to KK}=
    A_{\sigma_{l}\to KK}^{(1)}+
    A_{\sigma_{l}\to KK}^{(2)},
\eeq
where $A_{\sigma_{l}\to KK}^{(1)}$  originates from the
same source as $A_{\sigma_{l}\to\pi\pi}^{(1)}$ and
is determined by the kaon mass:
\beq
A_{\sigma_{l}\to KK}^{(1)}=-\frac{2M_K^2}{\chi_c}b_{\chi'\sigma_{l}},
\label{amplitudeSigmaTo2K}
\eeq
while the other part $A_{\sigma_{l}\to KK}^{(2)}$ again comes
from quark loop diagrams (see Appendix A).
The decay widths  thereby are\footnote{
The decay of $\sigmaII$ into kaons occurs almost at the threshold,
therefore, we cannot give a reliable  estimate for this process.
}
\ba
\Decay{\sigmaIII\to KK}{}&\approx & 260 \mev (125 \mev),\nonumber\\
\Decay{\sigmaIV\to KK}{}&\approx & 28 \mev, \nonumber\\
\Decay{\sigmaV\to KK}{}&\approx & 250 \mev.
\ea
The state $\sigmaI$  cannot decay into
kaons, as it is  below the threshold.

The amplitude describing decays of scalar isoscalar mesons
into $\eta\eta$ has a more complicated form, because
it contains a contribution from $\Delta L_{\rm an}$.
The singlet-octet mixing between pseudoscalar isoscalar
states should also be taken into account here.
Using the expression for the fields $\phi_{8,1}$ and
$\phi_{9,1}$ through the physical ones $\eta$ and $\eta'$:
\ba
\phi_{8,1}&=&b_{\phi_{8,1}\eta}\eta+b_{\phi_{8,1}\eta'}\eta'+\dots,\\
\phi_{9,1}&=&b_{\phi_{9,1}\eta}\eta+b_{\phi_{9,1}\eta'}\eta'+\dots,
\ea
where $\dots$ stand for the excited $\eta$ and $\eta'$ that
we do not need here and therefore omit them.
The mixing coefficients  for the scalar pseudoscalar
meson states were calculated in \cite{IJMPA1999,YF,ECHAJA2000}.
In the current calculation their values changed little
because the parameter $K$ has changed,
thus (see Table~\ref{etas}), $b_{\phi_{8,1}\eta}= 0.777$,
$b_{\phi_{8,1}\eta'}= -0.359$,
$b_{\phi_{9,1}\eta}= 0.546$, $b_{\phi_{9,1}\eta'}= 0.701$.
Thus, we obtain
for the amplitude:
\ba
A_{\sigma_{l}\to\eta\eta}=
    A_{\sigma_{l}\to\eta\eta}^{(1)}+
    A_{\sigma_{l}\to\eta\eta}^{(2)}+
    A_{\sigma_{l}\to\eta\eta}^{(3)}.
\ea
Here the contact term $A_{\sigma_{l}\to\eta\eta}^{(1)}$
has the form:
\beq
A_{\sigma_{l}\to\eta\eta}^{(1)}=-\frac{M_{\eta}^2}{\chi_c}b_{\chi'\sigma_{l}}
\eeq
The second term $A_{\sigma_{l}\to\eta\eta}^{(2)}$ comes from
a quark loop calculation (see Appendix A), and the third term
$A_{\sigma_{l}\to\eta\eta}^{(3)}$ originates from
$\Delta L_{\rm an}$ (see (\ref{Deltaanomalynotscaled})):
\beq
A_{\sigma_{l}\to\eta\eta}^{(3)}=\frac{2\hps}{3\chi_c}
\left(\sqrt{2}b_{\phi_{8,1}\eta}-b_{\phi_{9,1}\eta}\right)^2.
\eeq
As result, we obtain the following decay widths:
\ba
\Decay{\sigmaIII\to\eta\eta}{}&\approx & 62 \mev (26 \mev),\nonumber\\
\Decay{\sigmaIV\to\eta\eta}{}&\approx & 4 \mev, \nonumber\\
\Decay{\sigmaV\to\eta\eta}{}&\approx & 23 \mev.
\ea

The state $\sigmaV$ can also decay into $\eta\eta'$.
The corresponding amplitude is
\ba
A_{\sigma_{l}\to\eta\eta'}=
    A_{\sigma_{l}\to\eta\eta'}^{(2)}+
    A_{\sigma_{l}\to\eta\eta'}^{(3)}.
\ea
The contact term $A_{\sigma_{l}\to\eta\eta'}^{(1)}$
is absent here. The term
$A_{\sigma_{l}\to\eta\eta'}^{(2)}$ comes from
quark loop diagrams, as usual, and the last term
has the form:
\beq
A_{\sigma_{l}\to\eta\eta'}^{(3)}=\frac{4\hps}{3\chi_c}
\left(\sqrt{2}b_{\phi_{8,1}\eta}-b_{\phi_{9,1}\eta}\right)
\!\left(\sqrt{2}b_{\phi_{8,1}\eta'}-b_{\phi_{9,1}\eta'}\right).
\eeq
The decay width is approximately equal to 100 MeV.

The scalar meson states $\sigmaIII$, $\sigmaIV$, and $\sigmaV$
can decay into four pions. This decay can occur via
intermediate scalar mesons.
Similar calculations for $f_0(1500)$ were done
in our previous works \cite{EPJA2000,YF2000}.
Insofar as our calculations are qualitative,
we consider here, instead of the direct processes that
involve $\sigmaI$-resonances, simpler decays: into
$2\sigmaI$ and $\sigmaI2\pi$ as  final states.
Our investigation have shown that the result thus obtained can be
a good estimate for the decay into $4\pi$.

Let us consider decays into $2\sigmaI$. Its amplitude
can be divided into two parts:
\beq
A_{\sigma_{l}\to\sigmaI\sigmaI}=
    A_{\sigma_{l}\to\sigmaI\sigmaI}^{(1)}+
    A_{\sigma_{l}\to\sigmaI\sigmaI}^{(2)}.
    \label{sigmato2sigma}
\eeq
To calculate the first term $A_{\sigma_{l}\to\sigmaI\sigmaI}^{(1)}$,
one should first take, from the effective meson Lagrangian,
the terms that contains only scalar meson fields in the
third degree before taking account of mixing effects.
The corresponding vertices have the form:
\beq
a_1{\chi'}^3+a_2{\chi'}^2\sigma_{8,1}+
a_3{\chi'}\sigma_{8,1}^2+a_4\chi'\sigma_{8,2}^2,
\label{contactterms}
\eeq
where the coefficients $a_k$ are given in
Appendix A (see (\ref{a1})--(\ref{a2})).
These vertices come from $\bar L_{\rm G}$, ${\cal L}(\chi)$,
and $\Delta L_{\rm an}$ (see eqs.~(\ref{LGr}), (\ref{chi})),
and (\ref{Deltaanomalynotscaled})).
We neglected here
the terms with $\sigma_{9,i}$ fields which
represent quarkonia made of $s$-quarks,
because we are interested in decays into pions
that do not contain $s$-quarks.

Up to this moment, the contribution
$A_{\sigma_{l}\to\sigmaI\sigmaI}^{(1)}$
was considered.
As to the term $A_{\sigma_{l}\to\sigmaI\sigmaI}^{(2)}$
in (\ref{sigmato2sigma}) connected with
quark loops, its calculation is given in Appendix.
As a result, we obtain the following decay widths:
\ba
\Decay{\sigmaIII\to\sigmaI\sigmaI}{}&\approx & 40 \mev, \nonumber\\
\Decay{\sigmaIV\to\sigmaI\sigmaI}{}&\approx & 200 \mev, \nonumber\\
\Decay{\sigmaV\to\sigmaI\sigmaI}{}&\approx & 1 \mev.
\ea

Four pions in the final state can be produced also through the
process with one $\sigmaI$-resonance $(\sigma_{l}\to\sigmaI2\pi\to4\pi)$.
To estimate this process, we calculate the decay into
$\sigma2\pi$ as a final state.
The amplitude again can be divided into two parts:
\beq
A_{\sigma_{l}\to\sigmaI2\pi}=
    A_{\sigma_{l}\to\sigmaI2\pi}^{(1)}+
    A_{\sigma_{l}\to\sigmaI2\pi}^{(2)}.
\eeq
The first term has the form:
\ba
&&A_{\sigma_{l}\to\sigmaI2\pi}^{(1)}=
    -\frac{M_\pi^2}{\chi_c^2} b_{\chi'\sigma_{l}}b_{\chi'\sigmaI} +
    \frac{8m_{\rm u}}{\chi_c}b_{\chi'\sigma_{l}}\JJ{2}{0}{\bar  f_{\sigmaI}\bar f_\pi\bar f_\pi}\nonumber\\
&&\quad+ \frac{8m_{\rm u}}{\chi_c}b_{\chi'\sigmaI}\JJ{2}{0}{\bar  f_{\sigma_{l}}\bar f_\pi\bar f_\pi},
\ea
where $\bar f_a$ are ``physical'' form factors defined in Appendix A.
The pure quark contribution is calculated as described in
Appendix A. The result is
\beq
A_{\sigma_{l}\to\sigmaI2\pi}^{(2)}=
-4\JJ{2}{0}{\bar  f_{\sigma_{l}}\bar f_{\sigmaI}\bar f_\pi\bar f_\pi}.
\eeq
The corresponding decay widths are negligibly  small
\ba
\Decay{\sigmaIII\to\sigma2\pi}{}&\approx & 1 \mev, \nonumber\\
\Decay{\sigmaIV\to\sigma2\pi}{}&\approx & 2 \mev, \nonumber\\
\Decay{\sigmaV\to\sigma2\pi}{}&\approx & 0.6 \mev.
\ea

Comparing the obtained results with experimental data
(see Table \ref{Gdecays}), one can see that the decays $\sigmaI\to\pi\pi$
and $\sigmaII\to\pi\pi$ are in satisfactory agreement
with experiment.
For the states $\sigmaIII$, $\sigmaIV$, and $\sigmaV$,
we have reliable values only for their total widths
measured  experimentally. Our results allow us to obtain
just the order of magnitude for the decay widths, exceeding
the experimental values by a factor of $2.0\div 3.0$.

Concerning partial decay modes, the state $f_0(1500)$
decays mostly into $4\pi$ and $2\pi$. According to
the experimental data analysis given in
\cite{WA102}, the ratio $\Gamma_{4\pi}/\Gamma_{2\pi}\approx 1.34$.
We obtain $\Gamma_{4\pi}/\Gamma_{2\pi}\approx 2$, which is
in qualitative agreement with \cite{WA102}.
The decays into $4\pi$ and $2\pi$ are suppressed
for the state $f_0(1710)$.
Its main decay mode is into kaons.
This agrees with the analysis of
 experimental data given in \cite{WA102}
and corroborates our assumption that $f_0(1500)$
is rather a glueball.

\section{Conclusion and discussion}
In papers \cite{IJMPA1999,YF,ECHAJA2000}, we have shown
for the first time that  18 scalar  meson states
with masses lying between 0.4 GeV and 1.7 GeV can be considered
as two nonets of scalar quarkonia.  However,
in the mass interval under consideration,
there is an additional
meson state  which  is used to  be associated
with a scalar glueball. Two experimentally
observed scalars, $f_0(1500)$ and $f_0(1710)$,
are argued to be the most
probable candidates for
the glueball \cite{Anisovich,Jaminon,other}.
In \cite{IJMPA1999,YF,ECHAJA2000},
we have shown that the state $f_0(1710)$ is rather a quarkonium.
This conclusion was based on the analysis of strong decays of
both $f_0(1500)$ and $f_0(1710)$, assuming that one
of them is a glueball, and the other is a quarkonium.
The final decision
should be made after introducing the glueball into the
effective meson Lagrangian.

A chiral quark model for the description of
the ground state nonet only and
the scalar glueball
was suggested in \cite{Acta,EPJA2000,YF2000}.
There, our assumption that $f_0(1500)$ is the glueball
was corroborated. In the present work,
we extended the model \cite{EPJA2000,YF2000}
by introducing first radially excited states.
As a result, we obtained the complete description of
all 19 scalar mesons in the mass interval
concerned\footnote{
In the present work, only isoscalar mesons and the scalar glueball
were considered. As to the isovector and
strange mesons, they were studied in \cite{IJMPA1999,YF,ECHAJA2000},
and the introduction of the glueball has had
small effect on them.
}.

The basic parameters of the model $\Lambda$, $G$, $m_{\rm u}$, and $m_{\rm s}$
did not change either  after the introduction of the radially
excited states nor after the introduction of the glueball.
However, the parameter $K$ that describes the singlet-octet
mixing somewhat decreased, in  comparison with the
value used in \cite{IJMPA1999,YF,ECHAJA2000}, because
here, while fitting, we have taken into account
not only the masses of $\eta$ and $\eta'$ mesons
but also the lower experimental bound for the mass of $\sigmaI$
(400 MeV).
Let us emphasize  that due to the chiral symmetry, the form factor
parameters of scalar mesons are not arbitrary,
they coincide with those of pseudoscalar mesons.

In our model, we considered five scalar isoscalar meson
states: $\sigmaI$, $\sigmaII$, $\sigmaIII$, $\sigmaIV$,
and $\sigmaV$ with the masses: 400, 1070, 1320, 1550, and
1670 MeV, respectively. We identify them with
physically observed meson states in the following  sequence:
$f_0(400-1200)$, $f_0(980)$, $f_0(1370)$, $f_0(1500)$,
$f_0(1710)$ (see Table \ref{T:spectr}).
Note that, after the glueball is introduced into
the effective meson Lagrangian, the mass of $\sigmaI$
noticeably decreased as compared with the result from
\cite{IJMPA1999,YF,ECHAJA2000}.
This is a consequence of the noticeable mixing between
the glueball and the ground and radially excited
$\bar uu$ ($\bar dd$) quarkonia.
The obtained mass and decay width of $\sigmaI$ are in
satisfactory agreement with recent experimental data \cite{PDG,Li,Janssen}.
On the other hand, the $\bar ss$ quarkonia mix
with the glueball at a small proportion (see Table~\ref{mix}).
Therefore, after introducing the glueball (see \cite{IJMPA1999,YF,ECHAJA2000}),
the masses  of $\sigmaII$ and $\sigmaV$ change less
than the mass of $\sigmaI$.
However, here we obtain better agreement with experiment
for the mass of $\sigmaV$ than in \cite{IJMPA1999,YF,ECHAJA2000}.

The analysis of strong decay modes of the mesons mentioned above,
fulfilled in the framework of our investigation, corroborates
our former conclusion that the state $f_0(1710)$ is a quarkonium, while
$f_0(1500)$ consists mostly of the glueball.
Indeed, according to our calculations, the state
$f_0(1500)$ decays mostly into $4\pi$ and $2\pi$,
the decay into $4\pi$ being more probable. This is
in agreement with experiment \cite{PDG,WA102}.
Meanwhile, the decays of $f_0(1710)$ into $4\pi$ and $2\pi$
are suppressed as compared with those into kaons and $\eta$ mesons
(see \cite{PDG,WA102}).
On the other hand, if the model parameters were fixed
from the supposition that $f_0(1710)$ was the glueball, the
main decay mode of $f_0(1710)$ would be $4\pi$ ($\Gamma_{4\pi}=$150 MeV),
the remaining partial widths would be
too small: $\Gamma_{\pi\pi}=$3 MeV,
$\Gamma_{KK}=$5 MeV, $\Gamma_{\eta\eta}=$2 MeV,
$\Gamma_{\eta\eta'}=$2 MeV.
For the state $f_0(1500)$ in this case,
the main decay would be into kaons ($\Gamma_{KK}=$ 250 MeV),
the other modes would give: $\Gamma_{\pi\pi}$=10 MeV,
$\Gamma_{\eta\eta}=$ 34 MeV, $\Gamma_{4\pi}=$90 MeV.
This crucially disagrees with experiment \cite{WA102}.

For $\sigmaIV$, we obtain that state
contains 67\% of the glueball, which is in agreement
with \cite{Anisovich}.

Note that the decay of a scalar isoscalar meson into
four pions could go through a pair of $\rho$ mesons.
We tried to give an estimate of the decay width
for such a process in \cite{EPJA2000,YF2000}.
However, in the present paper, we did not consider
this process for the following reasons:
i) The interaction of the $\rho$-meson with the glueball
is beyond the model we considered here.
ii) There are specific problems
connected with gauge invariance.
A more thorough investigation is necessary  for an accurate solution
of the problem.

Let us remind that our model is based on the $U(3)\times U(3)$
chiral symmetry and scale invariance of an effective meson
Lagrangian.
Both symmetries are very approximate for the energies
under consideration. Therefore, our results are rather
qualitative. Nevertheless, we hope that the model gives,
on the whole,
a correct description of scalar meson properties.

\section*{Acknowledgement}
We thank
Drs.~S.B.~Gerasimov and A.E.~Dorokhov for useful discussions.
The work is supported by RFBR Grant 00-02-17190 and the
Heisenberg-Landau program 2001.
\bigskip

\appendix
\makeatletter
\@addtoreset{equation}{section}
\makeatother
\renewcommand{\theequation}{\thesection \arabic{equation}}

\section{Calculation of the quark loop contribution into the
strong decay amplitudes}
In the calculation of the quark loop contributions to
decay amplitudes,
we follow our papers \cite{IJMPA1999,YF,ECHAJA2000}, where
the external momentum dependence of decay amplitudes was taken into
account.

It is convenient to take account of the
mixing effects before integration.
To demonstrate how to do this,
let us first calculate the decay of the state $\sigmaI$ into
pions. As one can see, eight\footnote{
Two of them are identical, which leads to the  symmetry
factor of 2.
}
diagrams (Fig.~\ref{trigs}) contribute to this
process. The expression for the amplitude is as follows
(see (\ref{formfactors}) for the definition of form factors):
\ba
&&A_{\sigmaI\to\pi\pi}^{(2)}=
8m_{\rm u}\bigl[g_{8,1}b_{\sigma_{8,1}\sigmaI}
(g_{1,1}^2 Z b_{\pi_1\pi}^2\JJ{2}{0}{1}\nonumber\\
&&\quad+2g_{1,1}g_{1,2} \sqrt{Z} b_{\pi_1\pi}b_{\pi_2\pi}\JJ{2}{0}{f_2^1}+
g_{1,2}^2 b_{\pi_2\pi}^2\JJ{2}{0}{f_2^1 f_2^1})\nonumber\\
&&\quad+g_{8,2}b_{\sigma_{8,2}\sigmaI}(g_{1,1}^2 Z b_{\pi_1\pi}^2\JJ{2}{0}{ f_2^8}\nonumber\\
&&\quad+2g_{1,1}g_{1,2} \sqrt{Z} b_{\pi_1\pi}b_{\pi_2\pi}\JJ{2}{0}{f_2^8 f_2^1}\nonumber\\
&&\quad+g_{1,2}^2 b_{\pi_2\pi}^2\JJ{2}{0}{f_2^8f_2^1 f_2^1})\nonumber\\
&&\quad-P_1\cdot P_2\, (g_{8,1}b_{\sigma_{8,1}\sigmaI}
(g_{1,1}^2 Z b_{\pi_1\pi}^2\JJ{3}{0}{1}\nonumber\\
&&\quad+2g_{1,1}g_{1,2} \sqrt{Z} b_{\pi_1\pi}b_{\pi_2\pi}\JJ{3}{0}{f_2^1}+
g_{1,2}^2 b_{\pi_2\pi}^2\JJ{3}{0}{f_2^1 f_2^1})\nonumber\\
&&\quad+g_{8,2}b_{\sigma_{8,2}\sigmaI}(g_{1,1}^2 Z b_{\pi_1\pi}^2\JJ{3}{0}{f_2^8}\nonumber\\
&&\quad+2g_{1,1}g_{1,2} \sqrt{Z} b_{\pi_1\pi}b_{\pi_2\pi}\JJ{3}{0}{f_2^8 f_2^1}\nonumber\\
&&\quad+g_{1,2}^2 b_{\pi_2\pi}^2\JJ{3}{0}{f_2^8 f_2^1 f_2^1}))\bigr],
\label{Asigmapipibare}
\ea
where
\beq
f^1_2=c_\pi(1+d_{\rm u} \vec{k}^2);  \quad f^{8}_2=c_{8}(1+d_{\rm u}\vec{k}^2)
\eeq
and $c_\pi\equiv c_1=c_2=c_3=1.39$. The coefficient
$c_\pi$ is fixed by the mass of the radially excited
pion $\pi(1300)$.  This is described in
\cite{IJMPA1999,YF,ECHAJA2000,VolkovWeiss,Volk1997}.
The coefficient $c_8$ is given in (\ref{c8c9}).
The product of the momenta of  secondary
particles can be expressed through masses of mesons:
\beq
P_1\cdot P_2=\frac12(M^2-M_1^2-M_2^2),
\eeq
where $M$ is the mass of the decaying meson, and
$M_1$ and $M_2$ are the masses of secondary particles
($M=M_{\sigmaI}$, $M_1=M_2=M_\pi$ in this case).
Let us continue (\ref{Asigmapipibare}) and
calculate the sum before integration. The resulting expression
becomes short:
\ba
&&A_{\sigmaI\to\pi\pi}^{(2)}=
    8 m_{\rm u} (\JJ{2}{0}{\bar f_{\sigmaI}\bar f_\pi\bar f_\pi}\nonumber\\
    &&\quad-P_1\cdot P_2 \JJ{3}{0}{\bar f_{\sigmaI}\bar f_\pi\bar f_\pi}),
\ea
where $\bar f_a$ are  form factors for the physical
meson states, defined as follows:
\beq
\bar f_{\sigmaI}=g_{8,1}b_{\sigma_{8,1}\sigmaI}+g_{8,2}b_{\sigma_{8,2}\sigmaI}f^8_2,
\eeq
\beq
\bar f_\pi=g_{1,1}b_{\pi_{1}\pi}\sqrt{Z}+g_{1,2}b_{\pi_2\pi}f^1_2.
\eeq
The coefficients $b_{{\pi_1}\pi}$ appear because of the mixing
between the ground and excited pion states.
Their values are: $b_{\pi_1\pi}\approx 0.997$,
$b_{\pi_2\pi}\approx 0.007$  (see Table~\ref{pions}).
Concerning the decays into the other pairs of pseudoscalars,
the calculation of the corresponding contribution is
carried out in the same manner.
We will discriminate these form factors by the superscripts
$u$ and $s$, respectively. Below we give the physical
form factors that were used in the calculation:
\ba
&&\bar f_{\sigma_{l}}^{\rm u}=
    g_{8,1}b_{\sigma_{8,1}\sigma_{l}}
    +g_{8,2}c_{8} (1+d_{\rm u} \vec{k}^2) b_{\sigma_{8,2}\sigma_{l}},\\
&&\bar f_{\sigma_{l}}^{\rm s}=
    g_{9,1}b_{\sigma_{9,1}\sigma_{l}}
    +g_{9,2}c_{9} (1+d_{\rm s} \vec{k}^2) b_{\sigma_{9,2}\sigma_{l}},\\
&&\bar f_{\pi}=g_{1,1}b_{\pi_1\pi}\sqrt{Z}
    +g_{1,2}c_\pi (1+d_{\rm u} \vec{k}^2) b_{\pi_2\pi}, \\
&&\bar f_{K}=g_{4,1}b_{K_1K}\sqrt{Z}
    +g_{4,2}c_K (1+d_{\rm us} \vec{k}^2) b_{K_2K}, \\
&&\bar f_{\eta}^{\rm u}= g_{8,1}b_{\phi_{8,1}\eta}\sqrt{Z}
    +g_{8,2}c_{8} (1+d_{\rm u} \vec{k}^2) b_{\phi_{8,2}\eta},\\
&&\bar f_{\eta'}^{\rm u}= g_{8,1}b_{\phi_{8,1}\eta'}\sqrt{Z}
    +g_{8,2}c_{8} (1+d_{\rm u} \vec{k}^2) b_{\phi_{8,2}\eta'},\\
&&\bar f_{\eta}^{\rm s}= g_{9,1}b_{\phi_{9,1}\eta}\sqrt{Z}
    +g_{9,2}c_{9} (1+d_{\rm s} \vec{k}^2) b_{\phi_{9,2}\eta},\\
&&\bar f_{\eta'}^{\rm s}= g_{9,1}b_{\phi_{9,1}\eta'}\sqrt{Z}
    +g_{9,2}c_{9} (1+d_{\rm s} \vec{k}^2) b_{\phi_{9,2}\eta'}.
\ea
To calculate these form factors,
one needs, besides $c_\pi$, also $c_K\equiv c_4=c_5=c_6=c_7=1.6$
fixed by the mass of $K^*_0(1430)$,
and $d_{\rm us}=-1.75$ GeV$^{-2}$
(see \cite{IJMPA1999,YF,ECHAJA2000,Volk1997}).
The last parameter is fixed by a condition similar
to the ones that determine the parameters $d_{\rm u}$ and $d_{\rm s}$:
\beq
\JJ{1}{0}{1+d_{\rm us}\vec{k}^2}+
\JJ{0}{1}{1+d_{\rm us}\vec{k}^2}=0.
\eeq
The mixing coefficients for pseudoscalar mesons are given
in Tables~\ref{pions}--\ref{etas}.

Let us write the quark-loop contribution to the vertices
of the effective meson Lagrangian in terms of physical meson
states. Only the vertices describing the processes, which we
are interested in, are given below. For $l=$I,II,III,IV, V,
we have
\ba
&&A_{\sigma_{l}\to\pi\pi}^{(2)} \sigma_{l} (2\pi^+\pi^-\!+\pi^0\pi^0)\nonumber\\
&&+A_{\sigma_{l}\to KK}^{(2)} \sigma_{l} (K^+K^-\! + K^0\tilde K^0)\nonumber\\
&&+A_{\sigma_{l}\to \eta\eta}^{(2)} \sigma_{l}\eta\eta+
A_{\sigma_{l}\to \eta\eta'}^{(2)} \sigma_{l}\eta\eta'.
\ea

\ba
&&A_{\sigma_{l}\to\pi\pi}^{(2)}=
8 m_{\rm u} (\JJ{2}{0}{\bar f_{\sigma_{l}}^{\rm u}\bar f_\pi \bar f_\pi}
 -P_1\cdot P_2 \JJ{3}{0}{\bar f_{\sigma_{l}}^{\rm u}\bar f_\pi \bar f_\pi}),\nonumber\\
&&A_{\sigma_{l}\to KK}^{(2)}\!=\!
8 m_{\rm u} (C_{\rm uu} \JJ{2}{0}{\bar f_{\sigma_{l}}^{\rm u}\bar f_K \bar f_K}\!+\!
C_{\rm us}\JJ{1}{1}{\bar f_{\sigma_{l}}^{\rm u}\bar f_K \bar f_K}),\nonumber\\
&&\quad-
8\sqrt{2}m_{\rm s}(C_{\rm ss} \JJ{0}{2}{\bar f_{\sigma_{l}}^{\rm s}\bar f_K \bar f_K}+
C_{\rm su}\JJ{1}{1}{\bar f_{\sigma_{l}}^{\rm s}\bar f_K \bar f_K}),\nonumber\\
&&\quad-P_1\cdot P_2(8 m_{\rm s}\JJ{2}{1}{\bar f_{\sigma_{l}}^{\rm u}\bar f_K \bar f_K}
-8\sqrt{2}m_{\rm u} \JJ{1}{2}{\bar f_{\sigma_{l}}^{\rm s}\bar f_K \bar f_K}),\nonumber\\
&&A_{\sigma_{l}\to\eta\eta}^{(2)}=
8 m_{\rm u} \JJ{2}{0}{\bar f_{\sigma_{l}}^{\rm u}\bar f_\eta^{\rm u} \bar f_\eta^{\rm u}}
-8\sqrt{2} m_{\rm s} \JJ{0}{2}{\bar f_{\sigma_{l}}^{\rm s}\bar f_\eta^{\rm s} \bar f_\eta^{\rm s}},
\nonumber\\
&&\quad-P_1\cdot P_2 (8m_{\rm u}\JJ{3}{0}{\bar f_{\sigma_{l}}^{\rm u}\bar f_\eta^{\rm u} \bar f_\eta^{\rm u}}
-8\sqrt{2} m_{\rm s} \JJ{0}{3}{\bar f_{\sigma_{l}}^{\rm s}\bar f_\eta^{\rm s} \bar f_\eta^{\rm s}}),\nonumber\\
&&A_{\sigma_{l}\to\eta\eta'}^{(2)}=16 m_{\rm u} \JJ{2}{0}{\bar f_{\sigma_{l}}^{\rm u}\bar f_\eta^{\rm u} \bar f_{\eta'}^{\rm u}}\!-\!16
\sqrt{2} m_{\rm s} \JJ{0}{2}{\bar f_{\sigma_{l}}^{\rm s}\bar f_\eta^{\rm s} \bar f_{\eta'}^{\rm s}},\nonumber\\
&&\quad-P_1\cdot P_2 (16m_{\rm u}\JJ{3}{0}{\bar f_{\sigma_{l}}^{\rm u}\bar f_\eta^{\rm u} \bar f_{\eta'}^{\rm u}}
-16\sqrt{2} m_{\rm s} \JJ{0}{3}{\bar f_{\sigma_{l}}^{\rm s}\bar f_\eta^{\rm s} \bar f_{\eta'}^{\rm s}}),\nonumber\\
\ea
where
\ba
&&C_{\rm uu}=\frac{2m_{\rm u}}{m_{\rm u}+m_{\rm s}},\quad
C_{\rm us}=\frac{m_{\rm s}(m_{\rm u}-m_{\rm s})}{m_{\rm u}(m_{\rm u}+m_{\rm s})},\nonumber\\
&&C_{\rm ss}=\frac{2m_{\rm s}}{m_{\rm u}+m_{\rm s}},\quad
C_{\rm su}=\frac{m_{\rm u}(m_{\rm s}-m_{\rm u})}{m_{\rm s}(m_{\rm u}+m_{\rm s})}.
\ea

Now we consider the decays of a scalar isoscalar meson into
a pair of $\sigmaI$. To calculate the quark loop contribution
to the corresponding  decay amplitudes, one should follow
the method described above for the pseudoscalar mesons.
The quark loop contribution can be represented as a set
of diagrams that results in  a sum of integrals which
then can be converted into a single integral over
the physical form factors for scalar isoscalar mesons.
Thus, one obtains:
\beq
A_{\sigma_{l}\to\sigmaI\sigmaI}^{(2)}\approx 8m_{\rm u} \JJ{2}{0}{\bar f_{\sigma_{l}}^{\rm u}\bar f_{\sigmaI}^{\rm u}\bar f_{\sigmaI}^{\rm u}}
\eeq
for $l=$III, IV, V.
In conclusion, we display the coefficients $a_k$ that determine
contact terms (\ref{contactterms}):
\ba
a_1&=&-\frac{1}{\chi_c^3}
    \biggl[\frac{10}{3} \Cg+\sum_{a,b=8}^{9}\Bigl(-\frac{4}{3}\bar\mu^0_a(G^{(-)})^{-1}_{a b}\mu_b\nonumber\\
    &&+\frac{7}{3}\bar\mu^0_a\left(G^{(-)}\right)^{-1}_{a b}\bar\mu_b^0+
    \frac{1}{6}\mu^0_a\left(G^{(-)}\right)^{-1}_{a b}(\mu_b-\bar\mu^0_b)\Bigr)\nonumber\\
    &&+\hsc\bigl(16F_0^2-18F_0F_0^0+4(F_0^0)^2\bigr)\biggr],\label{a1}\\
a_2&=&
    -\frac{\sqrt{2}\hsc}{\sqrt{3}\chi_c^2}(14F_0-10F_0^0)\nonumber\\
    &&\quad-\frac{1}{\chi_c^2}\sum_{a=8}^9
    g_{8,1}\left(G^{(-)}\right)^{-1}_{8 a}\bar\mu_{a}^0,\\
a_3&=&
    \frac{4\hsc}{3\chi_c}\nonumber\\
    &&-\frac{1}{\chi_c}
    \left(g_{8,1}^2\!\left((G^{(-)})^{-1}_{8 8}\!-
    8I_1^{\Lambda}(m_{\rm u})\right)\!+\!4m_{\rm u}^2\right)\!,
    \\
a_4&=&\frac{1}{\chi_c}
    \left(g_{8,2}^2\left(1/G-
    8\JJ{2}{0}{f_2^{\rm 8}f_2^{\rm 8}}\right)+4m_{\rm u}^2\right).
    \label{a2}
\ea

\section*{Figure captions}
\newlength{\secondcolumnwidth}
\setlength{\secondcolumnwidth}{\linewidth}
\addtolength{\secondcolumnwidth}{-3cm}
\begin{tabular}{p{20mm} p{\secondcolumnwidth}}
\textbf{Figure} \ref{loops}&
The set of diagrams contributing to the effective meson
Lagrangian: (a) tadpoles, (b) quadratic terms,
(c) triangle diagrams, and (d) boxes.\\[2mm]
\textbf{Figure} \ref{trigs}&
The set of diagrams describing the decay of a
scalar meson into a pair of pions.
The vertices where a form factor
occurs are marked by $\mathbf{f}$.
\end{tabular}

\clearpage

\begin{figure}[H]
\includegraphics[scale=0.8]{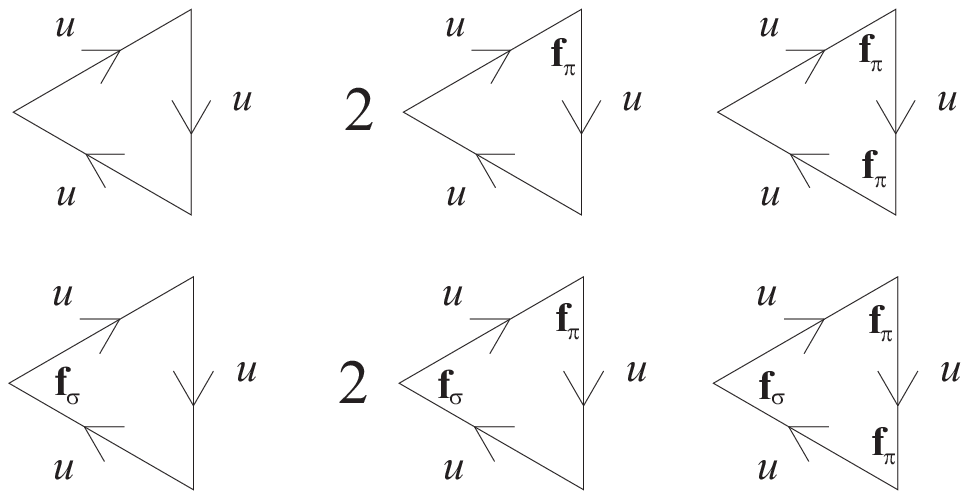}
\caption{}\label{trigs}
\end{figure}

\clearpage
\section*{Table captions}

\begin{tabular}{p{20mm}p{\secondcolumnwidth}}
\textbf{Table \ref{mix}.}& Elements of the matrix $b$,
describing mixing in the scalar isoscalar sector.
The singlet-octet and quarkonia-glueball mixing effects are
taken into account.\\[2mm]
\textbf{Table \ref{T:spectr}.}& The model and experimental masses
of scalar isoscalar meson states.\\[2mm]
\textbf{Table \ref{Gdecays}.}&
The partial and total decay widths (in MeV)
of scalar isoscalar meson states.
($^*$)For the meson state $\sigmaII$, there is
possible a decay into kaons, which we did not calculate
here, because its mass is at the threshold.
We show only the lowest limit for its total
decay width allowing for the decay into kaons
that can increase the total decay width.
The value is given for the model mass 1070 MeV.
Next, in brackets, we also give the decay width
calculated for the experimental mass 980 MeV.
In the case of $\sigmaIII$, two values are given
for its model mass and (in brackets) for the
lowest bound for its experimental mass (1200 MeV).
\\[2mm]
\textbf{Table \ref{pions}.}&
Mixing coefficients for the ground and excited  pion states.\\[2mm]
\textbf{Table \ref{kaons}.}&
Mixing coefficients for the ground and excited  kaon states.\\[2mm]
\textbf{Table \ref{etas}.}&
Mixing coefficients for the ground and excited  $\eta$ and $\eta'$ states.
Here, the singlet-octet mixing is taken into account.
\end{tabular}

\clearpage

\begin{table}[H]
\section*{Tables}
\centering
\begin{tabular}{||c|rrrrr||}
\hline
 & $\sigmaI$ & $\sigmaII$ & $\sigmaIII$ & $\sigmaIV$ & $\sigmaV$ \\
\hline
$\sigma_{8,1}$ & 0.973 & 0.137 & 0.393 & 0.548 & 0.048 \\
$\sigma_{8,2}$ & $-0.064$ & 0.204 & $-0.978$& $-0.647$ & $-0.047$ \\
$\sigma_{9,1}$ & $-0.225$ & 0.876 &  0.160 & 0.011 & 0.628 \\
$\sigma_{9,2}$ & $0.025$ & 0.146 &  0.136 & $-0.082$ & $-1.09$ \\
$\chi'$ & $-0.266$ & 0.095 &  $-0.495$ & $0.813$ & $-0.116$\\
\hline
\end{tabular}
\caption{}
\label{mix}
\end{table}

\begin{table}[H]
\centering
\begin{tabular}{||r|c|c||}
\hline
& Theor. & Exp. \cite{PDG}\\
\hline
$\sigmaI$ & 400 &  408 \cite{Li}, 387 \cite{Janssen}\\
$\sigmaII $ & 1070 & 980$\pm$10\\
$\sigmaIII$ & 1320 & 1200--1500\\
$\sigmaIV$ & 1550 & 1500$\pm$10\\
$\sigmaV$ & 1670 & 1712$\pm$5\\
\hline
\end{tabular}
\caption{}
\label{T:spectr}

\end{table}

\begin{table}[H]
\centering\begin{tabular}{||l|r|r|r|r|r||}
\hline
& $f_0(400-1200)$ & $f_0(980)$ & $f_0(1370)$ & $f_0(1500)$ & $f_0(1710)$\\
\hline
$\pi\pi$ & 600 & 36 (20)& 680 (480) & 100 & 0.3\\
$K \bar K$ & -- &  -- & 260 (125) &28 & 250\\
$\eta\eta$ & -- & -- & 62 (26) & 4 & 20\\
$\eta\eta'$ & -- & -- & -- & -- & 100\\
$4\pi (2\sigmaI)$ & -- & -- &  40 & 200 & 1\\
$\Gamma_{\rm tot}$ & 600 & $>40 (>20)(^*)$
 & 1040 (670) & 330 & 370\\
$\Gamma_{\rm tot}^{\rm exp}$ & 600--1200 & 40--100& 200--500&112$\pm$10 & 133$\pm$14\\
\hline
\end{tabular}
\caption{}
\label{Gdecays}
\end{table}

\begin{table}[H]
\centering\begin{tabular}{||c|cc||}
  \hline
  &$\pi$&$\pi'$\\ \hline
  $\pi_1$&0.997 & 0.511 \\
  $\pi_2$&0.007 & $-$1.12 \\ \hline
\end{tabular}
\caption{}\label{pions}
\end{table}

\begin{table}[H]
\centering\begin{tabular}{||c|cc||}
  \hline
  &$K$& $K'$\\\hline
  $K_1$ &0.954 & 0.533 \\
  $K_2$ &0.102 & $-$1.09 \\ \hline
\end{tabular}
\caption{}\label{kaons}
\end{table}

\begin{table}[H]
\centering\begin{tabular}{||c|cccc||}
  \hline
   & $\eta$ & $\eta'$ & $\hat\eta$ & $\hat\eta'$ \\ \hline
  $\phi_{8,1}$ & 0.777 & $-0.359$ & 0.668 & 0.276 \\
  $\phi_{8,2}$ & 0.102 & $-0.330$ & $-1.03$ & $-0.274$ \\
  $\phi_{9,1}$ & 0.546 & 0.701 & $-0.010$ & $-0.602$ \\
  $\phi_{9,2}$ & 0.037 & 0.225 & $-0.333$ & 0.994 \\ \hline
\end{tabular}
\caption{}\label{etas}
\end{table}

\end{document}